%% file: TCL_Potential_v10.tex
\documentclass[onecolumn,journal]{IEEEtran}
\IEEEoverridecommandlockouts
\usepackage{amsbsy,epsfig,epsf,url}
\usepackage{graphicx}
\usepackage{amssymb,amsmath,dsfont,amsfonts}
\usepackage{epstopdf}
\usepackage{cite}
\usepackage{color}
\usepackage{boxedminipage}
\usepackage{listings}
\usepackage{subfigure}
\usepackage{caption2}
\usepackage{float}
\usepackage{cases}
\usepackage{multirow}

\usepackage{pdfsync,srcltx}
\usepackage{amsmath}
\usepackage{amsfonts}
\usepackage{amssymb}
\usepackage{graphicx,subfigure,psfrag}
\usepackage{url}
\usepackage[colornames]{xcolor}
\usepackage{algorithmic}
\usepackage{algorithm}

\usepackage{acronym}
\input{my_preamble}

\newcommand{\cut}[1]{}

\renewcommand{\baselinestretch}{1}

\allowdisplaybreaks

\title{Potentials and Economics of Residential Thermal Loads Providing Regulation Reserve}

\author{
He Hao, Borhan M. Sanandaji, Kameshwar Poolla, and Tyrone L. Vincent 
\thanks{H. Hao, B. M. Sanandaji, and K. Poolla are with the Department of Electrical Engineering and Computer Sciences, University of California, Berkeley, CA 94720 (hehao@berkeley.edu).}
\thanks{T. L. Vincent is with the Department of Electrical Engineering and Computer Science, Colorado School of Mines, Golden, CO 80401.}
}

\begin{document}

\maketitle

\begin{abstract}
Residential \acp{TCL} such as Air Conditioners (ACs), heat pumps, water heaters, and refrigerators have an enormous thermal storage potential for providing regulation reserve to the grid. In this paper, we study the potential resource and economic analysis of TCLs providing frequency regulation service. In particular, we show that the potential resource of \acp{TCL} in California is more than enough for both current and predicted near-future regulation requirements for the California power system. Moreover, we estimate the cost and revenue of \acp{TCL}, discuss the qualification requirements, recommended policy changes, and participation incentive methods, and compare \acp{TCL} with other energy storage technologies. We show that \acp{TCL} are potentially more cost-effective than other energy storage technologies such as flywheels, Li-ion, advanced lead acid, and Zinc Bromide batteries.  
\end{abstract}

\acresetall
\section{Introduction}\label{sec:intro}

The reliability and stability of the power grid requires continuous balance between supply and demand on a second-to-second basis, which otherwise will result in catastrophic consequences. Ancillary services such as frequency regulation and load following play an important role in achieving this power balance under normal operating conditions. While load following handles more predictable and slower changes in load, frequency regulation mitigates faster changes in system load and corrects unintended fluctuations in generation \cite{AS_Kirby}.  Frequency regulation has been traditionally provided by relatively fast-responding generators. However, the ramping rate of these generators is generally slower than that of the regulation signal, which results in poor power quality and high regulation procurement \cite{AS_Kirby, ERCOT}. The regulation requirement can be lowered if faster responding resources are available \cite{kema2009fastresponse}. It has been shown if \ac{CAISO} dispatched fast responding regulation resources, it could reduce its regulation procurement by as much as $40\%$ \cite{pnnl2008value}. This issue has been recognized in the power system community. The recently issued FERC orders 784 and 755 require considering the speed and accuracy of regulation resources when procuring regulation service.

In accordance to FERC order 755, CAISO has introduced a mileage product to provide compensation for faster and more accurate regulation resources \cite{CAISO_755}. Moreover, CAISO's definition of a Non-Generator Resource (NGR) with Regulation Energy Management (REM) allows NGRs with limited energy capacities, such as batteries and flywheels, to competitively bid in the regulation market. REM resources can bid to provide power based on their $15$-minute energy capacity into the day-ahead ancillary service market, and \ac{CAISO} will dispatch these resources so that their State of Charge (SoC) limits are respected \cite{CAISO_storage}. These regulatory developments have roused a growing interest in tapping the potentials of  fast-responding and accurate regulation resources.

In this paper, we show that \acp{TCL} have a great potential for providing fast regulation service, due to their large population size and the ability of being turned ON/OFF simultaneously. The proof of concept of using \acp{TCL} to provide regulation reserve and load following has been reported in \cite{lu2012evaluation, mathieu2013state, Hiskens, zhang2013aggregated, bashash2013modeling}. Other related work include study of commercial HVAC (Heating, Ventilation, and Air-Conditioning) systems, residential pool pumps, and electric vehicles to provide ancillary services to the grid \cite{lin2013low, HH_YL_AK_PB_SM_TSG:14, oldewurtel2013towards, meyn2014ancillary,haocharacterizing,barooahspectral, kempton2008test,AN_JT_AS_KP_PV_CDC:13}.  In our recent work~\cite{hehao2013generalized,HH_BS_KP_TV_TPS_2013}, we have shown that the aggregate flexibility offered by a collection of \acp{TCL} can be succinctly modeled as a generalized battery with dissipation. A similar work that models TCLs as a battery (without dissipation) is given in \cite{mathieu2013energy}. Moreover, we analytically characterized the power limits and energy capacity of this battery model in terms of the \ac{TCL} parameters and exogenous variables such as ambient temperature and user-specified set-points. Based on this battery model, in this paper we estimate the potential of \acp{TCL} in California for regulation service provision. We show that conservative estimate of the available power is larger than twice of the current maximum regulation procurement ($600$ MW). Additionally,  it is larger than the predicted maximum regulation requirement of \ac{CAISO} with $33\%$ of renewable penetration ($1.3$ GW)  \cite{helman2010resource}. Moreover, conservative estimate of the available energy capacity is much larger than the maximum energy requirement for regulation with both the $600$ MW and $1.3$ GW power procurements. The potential of \acp{TCL} in California is more than enough for provision of regulation service for now and the near future.

We further estimate the cost and revenue of \acp{TCL} for providing regulation service to the grid. Due to the stringent telemetry and metering requirements of \ac{CAISO}, the real-time power measurement of each individual \ac{TCL} is required to be reported to the ISO every 4 seconds. This requirement imposes a non-trivial cost on each unit to satisfy the qualification requirements. Moreover, CAISO currently requires the minimum resource size to be $0.5$ MW, and no aggregation of loads is allowed. We comment that these rules must be changed in order to allow an aggregator to profitably provide regulation service using TCLs in the California regulation market. We also estimate the cost of instrumentation to enable TCLs to provide regulation service, and recommend new policies to integrate power measurement, external control, and communication capabilities into appliance standards to reduce their capital cost. Additionally, we show that the annual revenue per \ac{TCL} is not very attractive if the total revenue is split evenly to each unit.  Therefore, a fair and attractive incentive method needs to be designed to encourage customer participation.  Moreover, we compare \acp{TCL} with other energy storage technologies that are suitable for frequency regulation. We show that \acp{TCL} are more competitive than other storage technologies such as flywheels, Li-ion, advanced lead acid, and Zinc Bromide batteries. However, large scale implementations need to be conducted to showcase the feasibility of this method.

The work of \cite{mathieu_revenue, HH_BS_KP_TV_ACC:2014,macdonald2012demand} are closely related to the present paper. In \cite{mathieu_revenue,HH_BS_KP_TV_ACC:2014}, the authors estimated the potential and revenue of \acp{TCL} for providing frequency regulation and/or load following services. Different from the work in \cite{mathieu_revenue,HH_BS_KP_TV_ACC:2014} that estimate the revenue based on ``pay-by-capacity" scheme in the regulation market, we estimate the potential of TCLs for regulation provision based on the ``pay-for-performance" scheme using historic data of \ac{CAISO}. In \cite{macdonald2012demand}, the authors reviewed the historic ancillary service price, market size, and discussed the ancillary service qualification requirements for various ISOs in the United States. In this paper,  we focus on the regulation market in California, and give more details on the qualification requirements in CAISO for regulation service provision, and recommend certain policy changes to enable TCLs to participate in the CAISO regulation market.

The remainder of the paper unfolds as follows. In Section \ref{sec:flexibility}, we present a method of characterizing the aggregate flexibility of \acp{TCL} using a generalized battery model. On the basis of this generalized battery model, we study in Section \ref{sec:potential} the potential and revenue of \acp{TCL} for regulation service provision in California. In Section \ref{sec:econo}, we estimate the capital cost of \acp{TCL} for regulation service provision, discuss the customer incentive methods and qualification requirements, and compare \acp{TCL} with other energy storage technologies. In Section \ref{sec:conclusion}, we give conclusions and future work, and recommend certain policy changes in order to enable \acp{TCL} to participate in the CAISO regulation market.

\begin{table*}[tb]
\caption{Typical parameter values for AC, heat pump, water heater and refrigerator  \cite{mathieu_revenue}.}
{\scriptsize
\label{tab:model_parameters}
\begin{center}
\begin{tabular}{ccccccc}
\hline
Parameter & Description & Unit & \emph{AC} &  \emph{Heat Pump} &  \emph{Water Heater} &  \emph{Refrigerator} \\
\hline 
$C$ & thermal capacitance  & kWh/$^{\circ}{\rm C}$ & $1.5-2.5$ &$1.5-2.5$ & $0.2-0.6$ &$0.4-0.8$\\
$R$ & thermal resistance  & $^{\circ}{\rm C}$/kW &  $1.5-2.5$ & $1.5-2.5$& $100-140$& $80-100$\\
$P_m$ & rated electrical power  & kW &   $4-7.2$& $4-7.2$ & $4-5$& $0.1-0.5$\\
$\eta$ & coefficient of performance  &  & $2.5$& $3.5$ & $1$ & $2$\\
$\theta_r$ & temperature setpoint  &$^{\circ}{\rm C}$ & $18-27$ &$15-24$ & $43-54$ & $1.7-3.3$\\
$\Delta$ & temperature deadband  & $^{\circ}{\rm C}$ & $0.125-0.5$ & $0.125-0.5$& $1-2$& $0.5-1$\\
$\theta_a$ & ambient temperature & $^{\circ}{\rm C}$ & variable & variable& $20$& $20$\\
\hline
\end{tabular}
\end{center}
}
\end{table*}

\section{Methods}
\label{sec:flexibility}
In this section, we present a method of characterizing the aggregate flexibility of a large collection of \acp{TCL}. The central idea is a generalized battery model, which provides a simple, compact, and meaningful representation of the flexibility offered by \acp{TCL}. This generalized battery model is the foundation for studying the potential and revenue of \acp{TCL} for regulation service provision in the California power system. 
\subsection{Individual Model of \acp{TCL}}
\label{sec:model}
In this paper, we consider a large collection of Thermostatically Controlled Loads (TCLs). The temperature dynamics of each \ac{TCL} are described by a standard hybrid-system model
\begin{align}
\label{eq:hybrid_model}
 \dot{\theta}(t) =
  \begin{cases} 
	 a(\theta_a-\theta(t)) - b P_m + w, & \text{ON state, } q(t) = 1,\\	
	 a(\theta_a-\theta(t))  + w, & \text{OFF state, } q(t) = 0,
\end{cases}
\end{align}
where $\theta$ is the \ac{TCL} temperature, $\theta_a$ is the ambient temperature, $P_m$ is the rated power, and  $a = \frac{1}{C R}$, $b = \frac{\eta}{C}$ are given in terms of the thermal capacitance $C$, thermal resistance $R$, and coefficient of performance $\eta$. The term $w$ accounts for external disturbances from occupancy, appliances, and so on.  Each \ac{TCL} has a temperature setpoint $\theta_r$ with a hysteretic ON/OFF local control within a temperature band $[\theta_r - \Delta, \theta_r + \Delta]$.  The operating state $q(t)$ evolves as
\begin{align*}
\lim_{\epsilon \rightarrow 0} q(t+\epsilon) =
\begin{cases} 
	q(t), & |\theta(t) - \theta_r| < \Delta, \\
	1-q(t), & |\theta(t) - \theta_r| = \Delta.
\end{cases}
\end{align*}
The parameters that specify this \ac{TCL} model are $\chi = (a,b, \theta_a, \theta_r, \Delta, P_m)$. We consider four types of \acp{TCL}: AC, heat pump, water heater and refrigerator. Table I describes the parameters and their typical values \cite{mathieu_revenue}. 

The average power consumed by a \ac{TCL} over a cycle is 
\begin{align}\label{eq:average_power}
	P_o =  \frac{P_m T_{\textrm{ON}}}{T_{\textrm{ON}}+T_{\textrm{OFF}}},
\end{align}
where $T_{\textrm{ON}}$ and  $T_{\textrm{OFF}}$ are respectively the time it spends in the ON and OFF states per ON/OFF cycle. For a large collection of \acp{TCL} that is uncoordinated, the steady-state power draw will be very close to the summation of their average power consumption, because at any time instant, any specific \ac{TCL} will be at a random point along its ON/OFF cycle. For a large collection of  \acp{TCL} indexed by $k$, the baseline (steady-state) power is given by
\begin{align*}
	n(t)=\sum_{k}  P^k_o,
\end{align*}
where the average power $P^k_o$ is given in \eqref{eq:average_power}. 
Additionally, the aggregated instantaneous power consumption is 
\begin{align*}
	P_{\textrm{agg}}(t)=\sum_{k}  q^k(t) P^k_m.
\end{align*}

\subsection{Generalized Battery Model of \acp{TCL}}\label{sec:battery}
To enable TCLs to provide regulation service, we can override their internal control $q^k(t)$'s, so that the aggregate instantaneous power consumption $P_{\textrm{agg}}(t)$ minus the baseline power $n(t)$ follow a dispatched regulation signal. For example, we can turn off some of the ON units to ``provide" power to the grid, or  turn on some of the OFF units to ``absorb" power from the grid. However, before tracking a regulation signal, we need to characterize all the dispatch signals that a collection of TCLs can successfully follow while respecting their temperature requirements. We define the power perturbation for each TCL as $e^k(t)=q^k(t) P_m^k -P_o^k$. During a cycle, each \ac{TCL} can accept power perturbation ($e^{k}(t)$) around its average power consumption ($P_{o}^{k}$) that will still meet user-specified comfort temperature bounds.
Define 
\[ \mathbb{E}^k = \left\{ e^k(t) \Big | \begin{matrix}  0 \leq P_{o}^{k} + e^{k}(t)  \leq P_{m}^{k} \\ P_o^k +e^k(t) \text{ maintains } |\theta^k(t) - \theta_r^k| \leq \Delta^k
 \end{matrix} \right\}. 
 \]
This set of power perturbations represents the flexibility of the $k^{th}$ TCL, which includes all its admissible power deviations with respect to baseline.
The \emph{aggregate flexibility} $\mathbb{U}$ of the collection of TCLs is defined as the Minkowski sum
\begin{align*}
\mathbb{U} = \sum_k \mathbb{E}^k.
\end{align*}

The above set-theoretic representation of the aggregate flexibility of \acp{TCL} is very abstract, and it is not portable for the system operator to integrate \acp{TCL} into the power system operation and control. In order to represent the aggregate flexibility $\mathbb{U}$ in a simple, intuitive, and meaningful way, we characterized the aggregate flexibility of TCLs in our previous work \cite{hehao2013generalized, HH_BS_KP_TV_TPS_2013} using a generalized battery model. 
\begin{definition}
A \emph{Generalized Battery Model} $\mathbb{B}$ is a set of signals $u(t)$ that satisfy
\begin{align*}
-n_- \leq u(t) \leq n_+, \quad \forall \ t>0, \qquad \qquad \\
\dot{x}(t) = - \alpha x(t) - u(t), \ x(0) = 0 \  \Rightarrow |x(t)| \leq \mathcal{C}, \ \ \forall \ t>0.
\end{align*} 
The model is specified by non-negative parameters $\phi = ( \mathcal{C}, n_-, n_+, \alpha)$,  and we write it compactly as $\mathbb{B}(\phi)$. $\hfill \square$
\label{def:battery}
\end{definition}

We can regard $u(t)$ as the power draw of the battery, $x(t)$ as its SoC, $n_-$/$n_+$ as its charge/discharge rate, and $\mathcal{C}$ as its energy capacity. This battery model provides a succinct and compact framework to characterize the aggregate power limits and energy capacity of a population of \acp{TCL}.  Additionally, a battery model is already supported by the regulation markets of system operators such as \ac{CAISO} and \ac{PJM} \cite{CAISO_storage, PJM_storage}. We showed the aggregate flexibility $\mathbb{U}$ could be represented by a generalized battery model. We next present the results of \cite{hehao2013generalized, HH_BS_KP_TV_TPS_2013} in Theorem \ref{thm:main_results}.

\begin{theorem}\label{thm:main_results}
Consider a collection of $N$ \emph{homogeneous} \acp{TCL} parameterized by $\chi = (a,b, \theta_a, \theta_r, \Delta, P_m)$.  Then, 
\begin{align}
 \mathbb{U} = \mathbb{B}( \mathcal{C}, n_-,n_+, \alpha), 
\end{align} 
where the battery parameters are given by
\begin{align*}
\mathcal{C}=N\Delta/b, \ n_-= N P_{o}, \ n_+=N (P_{m}-P_{o}), \ \alpha =a,
\end{align*}
and $P_{o} \approx a(\theta_{a}-\theta_{r})/b$. $\hfill \Box$
\end{theorem}

The above theorem implies that if a dispatched regulation signal for a collection of \acp{TCL} is within the  battery model that is associated with them (i.e., it is within both the power limits and energy capacity of $\mathbb{B}(\phi)$), then the collection of \acp{TCL} can track this signal. However, if the regulation signal is not in the battery model (it exceeds either the power limits or energy capacity of $\mathbb{B}(\phi)$), then the collection of TCLs cannot track this regulation signal. This battery characterization informs the system operator what regulation signal should be dispatched to \acp{TCL}, and how much power a collection of \acp{TCL} can provide in the regulation market.  The discussion on how to deal with heterogeneous \acp{TCL}, and consideration of short-cycling constraints on \ac{TCL} unit was reported in our previous work \cite{BS_HH_KP_TV_ACC:2013}. Interested readers are referred to \cite{BS_HH_KP_TV_ACC:2013} for more details.

\section{Results}
\label{sec:potential}
\subsection{Potential Resource of \acp{TCL} in California}
In this section, we first estimate the potential of \acp{TCL} in California using Theorem \ref{thm:main_results}.

Prior to October of 2009, the maximum amounts of upward and downward regulation the \ac{CAISO} procured were respectively about $375$ MW and $500$ MW \cite{CAISO_regulation}. More recently, the amount of regulation has increased; the maximum amounts of upward and downward regulation that \ac{CAISO} procured from June 2013 to May 2014 were respectively $500$ MW and $600$ MW. Additionally, the maximum procured upward and downward mileages were respectively $3.8$ GW and $4.9$ GW \cite{OASIS}.  In particular, the hourly minimum, average, and maximum in-market capacity and mileage procurements are depicted in Fig.  1. Furthermore, it has been predicted that if California achieved its $33\%$ of renewable penetration target by 2020, both the maximum upward and downward regulation capacity procurements would increase to $1.3$ GW \cite{helman2010resource}. 

\begin{figure}[tb]
\centering
\subfigure[Regulation Up]{\includegraphics[width=.49\columnwidth]{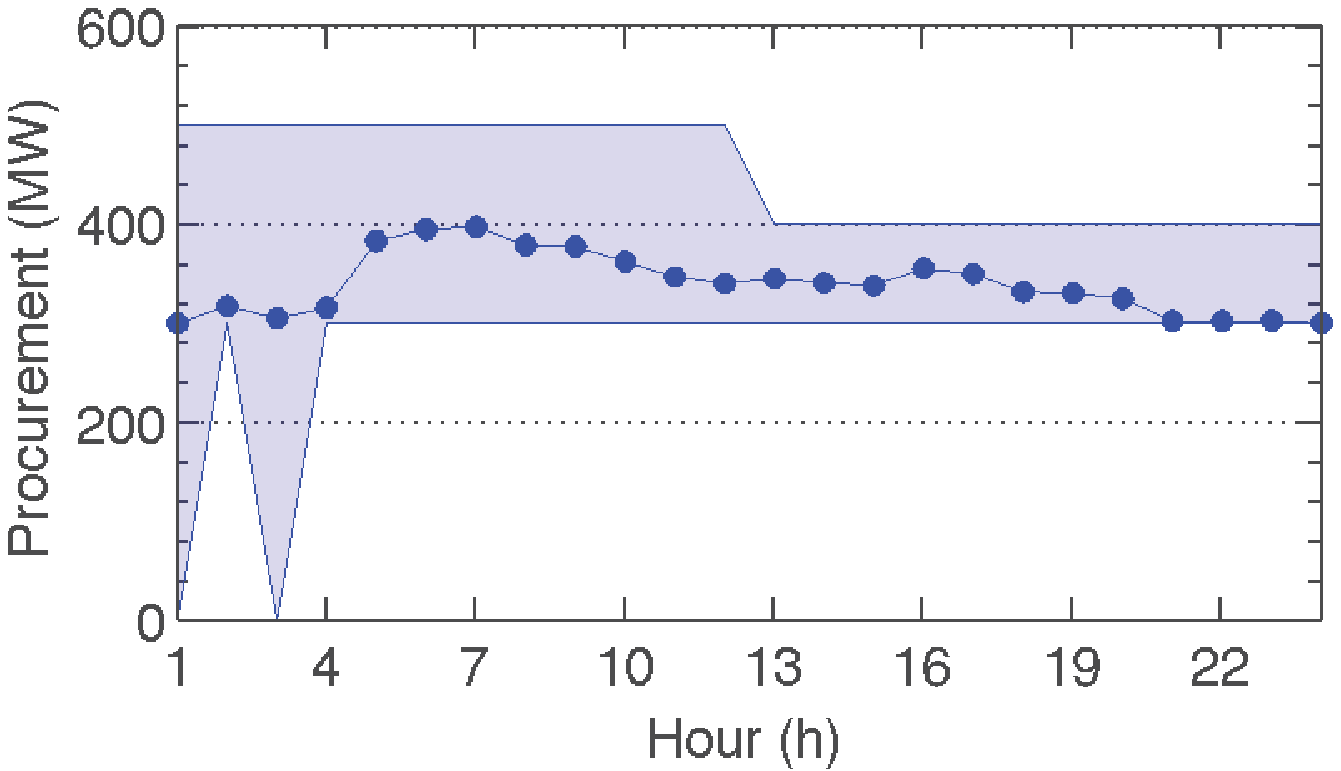}}
\subfigure[Regulation Down]{\includegraphics[width=.49\columnwidth]{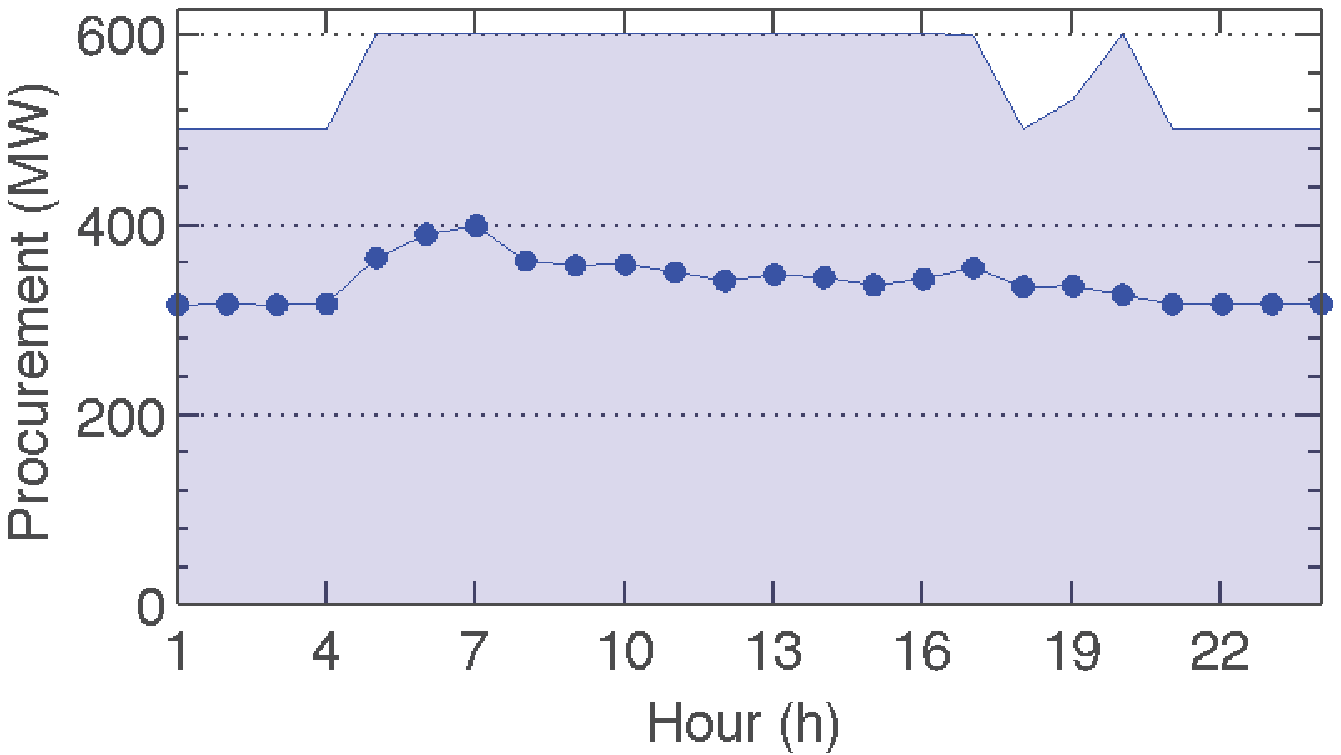}}
\subfigure[Regulation Mileage Up]{\includegraphics[width=.49\columnwidth]{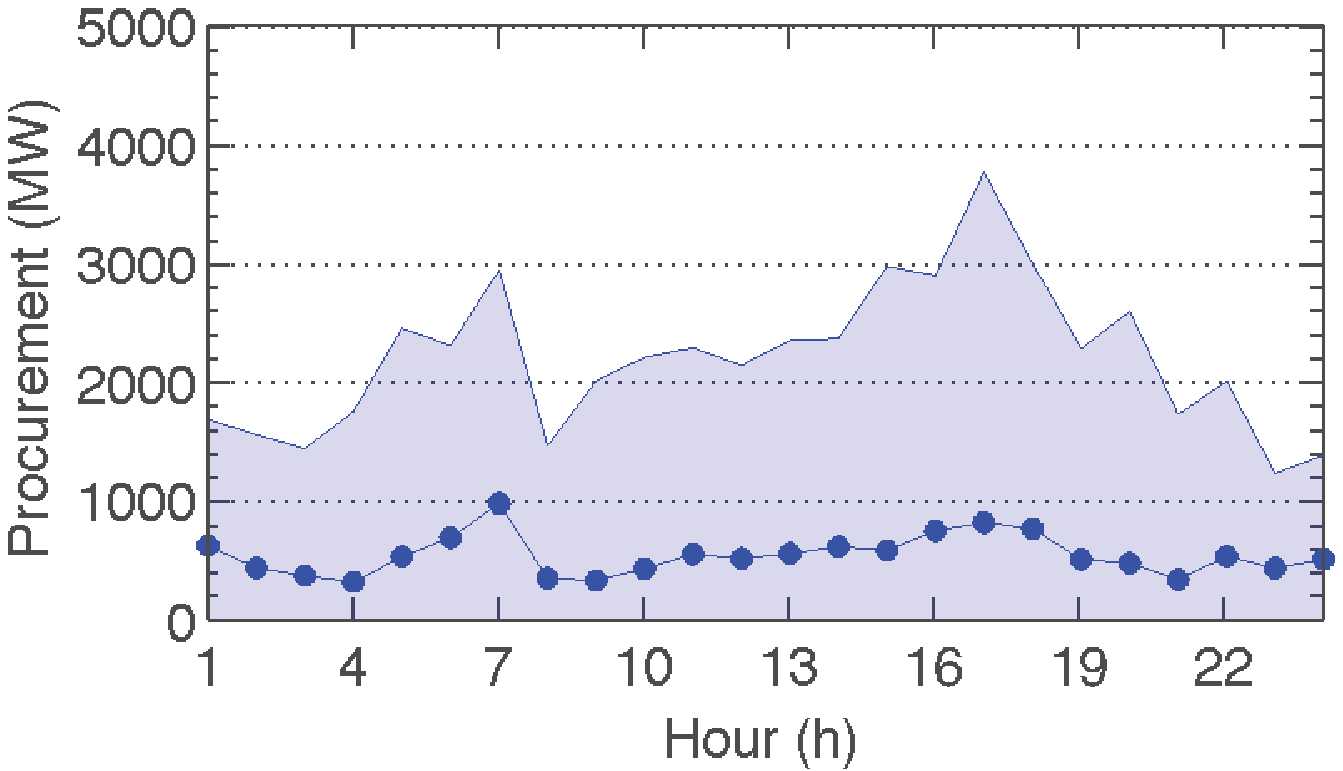}}
\subfigure[Regulation Mileage Down]{\includegraphics[width=.49\columnwidth]{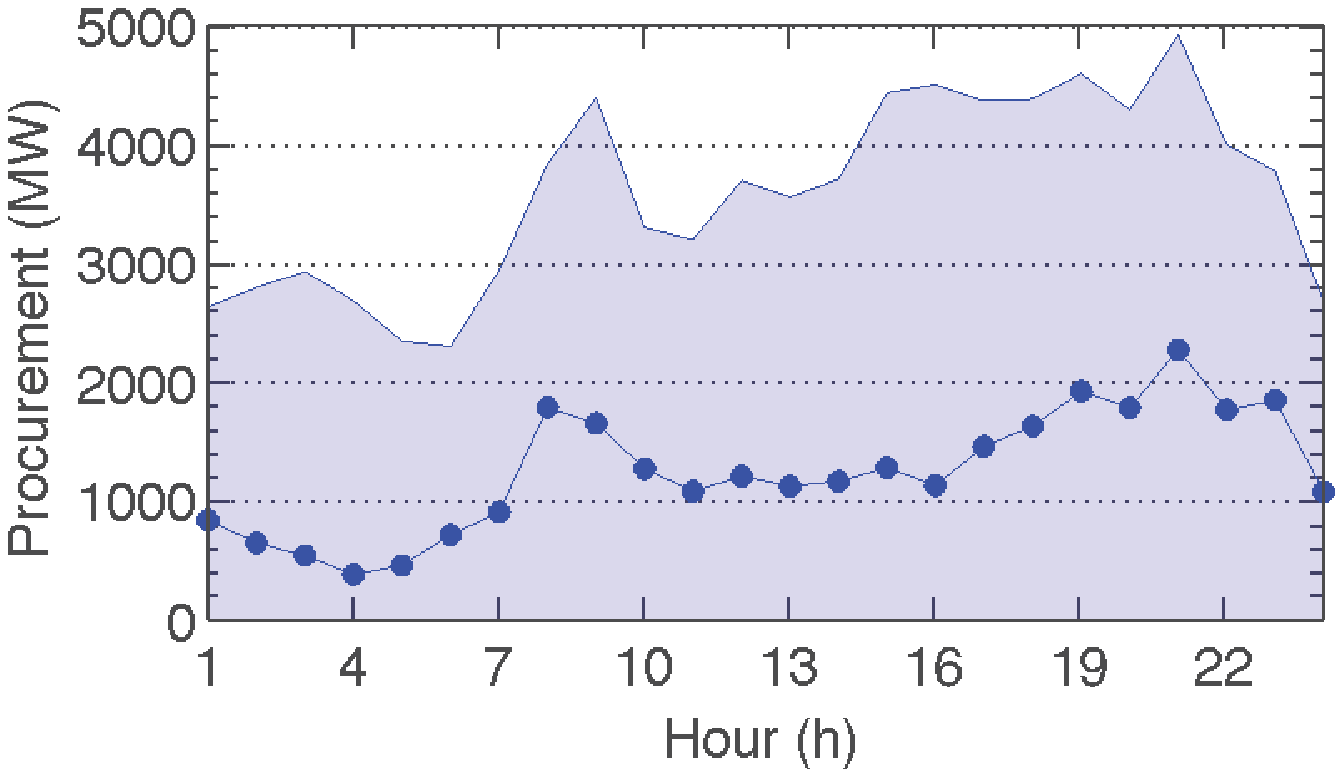}}
\caption{Hourly minimum, average, and maximum capacity procurements for upward and downward regulation and mileage in California. The plots are based on historic data of \ac{CAISO} from June 2013 to May 2014.}\label{fig:capacity}
\end{figure}

Using the generalized battery model, we estimate the available regulation capacity from the four types of \acp{TCL} listed in Table I. Note that the ambient  temperatures of water heater and refrigerator are assumed to be $20 ^\circ$C. This makes their power limits and energy capacities constant, independent of the outside temperature. This is not the case for other types of \acp{TCL} such as ACs and heat pumps. In addition, the percentage of participation of ACs and heat pumps is also a function of ambient  temperature. For example, in the case of ACs, there is less participation when the ambient temperature is low, and more participation when the ambient temperature is high. We make a heuristic assumption that participation percentages of ACs and heat pumps change over ambient temperature based on inverse tangent functions, which are depicted in Fig.  2. The examination of the precise participation functions requires extensive measurement data from the customer side, which is an important piece of future work.

\begin{figure}[tb]
\centering
\subfigure[ACs]{\includegraphics[width=.49\columnwidth]{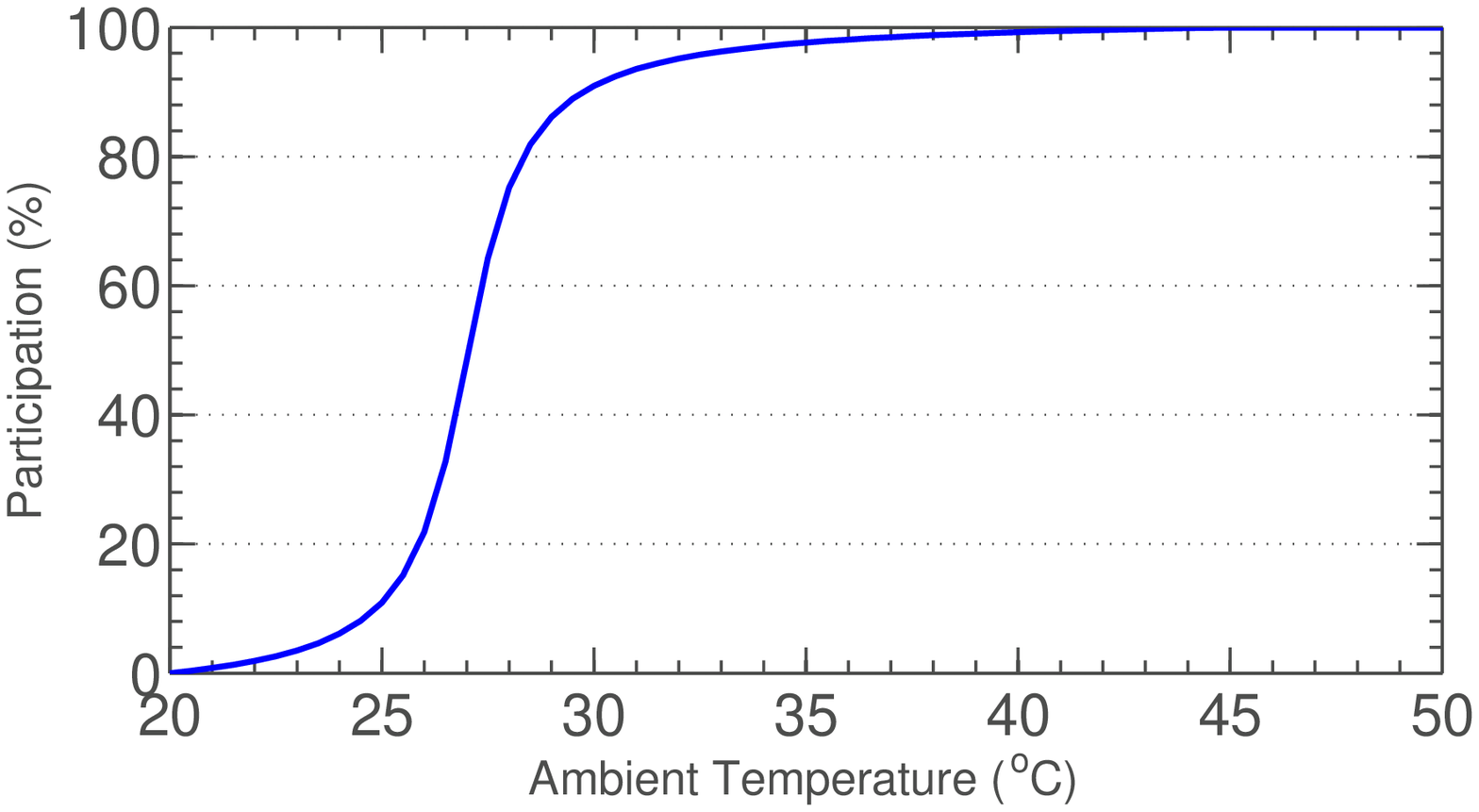}}
\subfigure[Heat Pumps]{\includegraphics[width=.49\columnwidth]{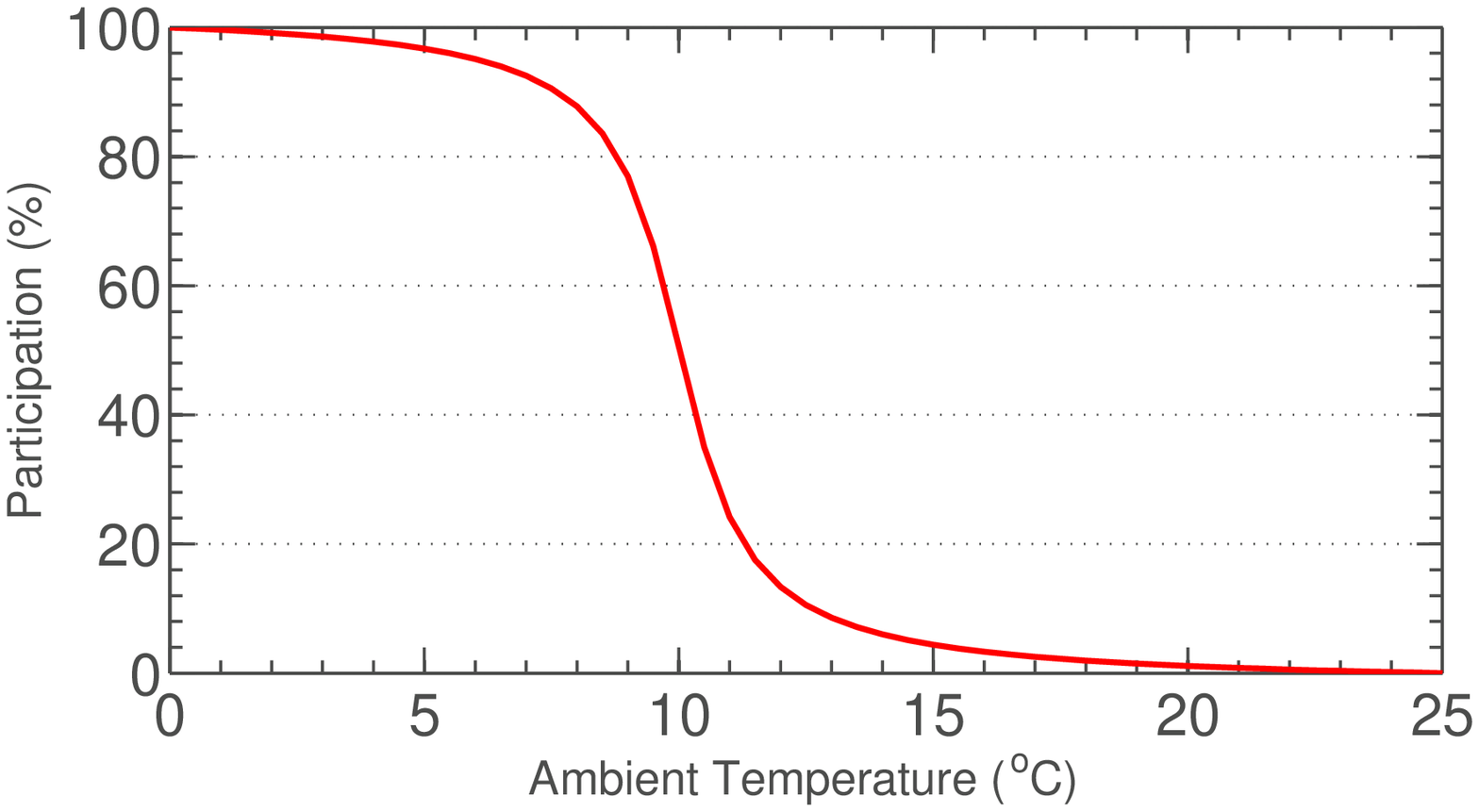}}
\caption{The functions of participation percentage over ambient temperature for ACs and heat pumps. } \label{fig:participation_vs_temp}
\end{figure}

Based on the census result in 2011, there are $13.7$ million households in California \cite{CA_census}. The saturation rates for the four types of \acp{TCL} are summarized in Table II based on a survey \cite{CA_applicance}. Note that the saturation rates for electrical heat pumps and water heaters are very low, this is because most of heat pumps and water heaters in California are using natural gas. Additionally, about half of the households possess ACs. For refrigerators,  the saturation rate is very high. On average, each house has 1.2 refrigerators. However, compared to other types of \acp{TCL}, the rated power of refrigerator is small, with an average of $0.3$ kW, while the rated power of the other \acp{TCL} is about $5$ kW (see Table I). 

\begin{table}[tb]
\caption{Saturation rates of AC, heat pump, water heater, and refrigerator in California.}
\label{tab:saturation}
{\scriptsize
\begin{center}
\begin{tabular}{c||c|c|c|c}
     & AC & Heat Pump &  Water Heater & Refrigerator \\
\hline \hline
{\scriptsize Percentage} & {\scriptsize $46.5 \%$} & {\scriptsize $1 \%$}& {\scriptsize$6.5\%$}  & {\scriptsize $122.3\%$} \\
{\scriptsize Number of units} & {\scriptsize $6.37\times10^6$} & {\scriptsize$0.14 \times10^6$}  & {\scriptsize $0.89\times10^6$} & {\scriptsize $16.75\times10^6$}\\
\end{tabular}
\end{center}
}
\end{table}

For each type of \ac{TCL}, we estimate their aggregate flexibility using parameter values as the mean of the values listed in Table I. We estimate their aggregate upward and downward power limits, and energy capacities for each type of \acp{TCL} using Theorem \ref{thm:main_results}, together with the corresponding participation functions of ACs and heat pumps given in Fig.  2.  Additionally, we assume the potential of TCLs in California can be represented by a weighted sum of the potentials of five cities: Sacramento (SA), San Francisco (SF), Bakersfield (BF), Los Angeles (LA), and San Diego (SD). The weight of each city is the same as the ratio of the number of households in their counties (SA 558,807, SF 380,971, BF 288,342, LA 3,462,202, and SD 1,176,718 \cite{CA_census}) to the total number of households in these five cities.

\begin{figure}[tb]
\centering
\subfigure[Hourly power limits of ACs]{\includegraphics[width=.49\columnwidth]{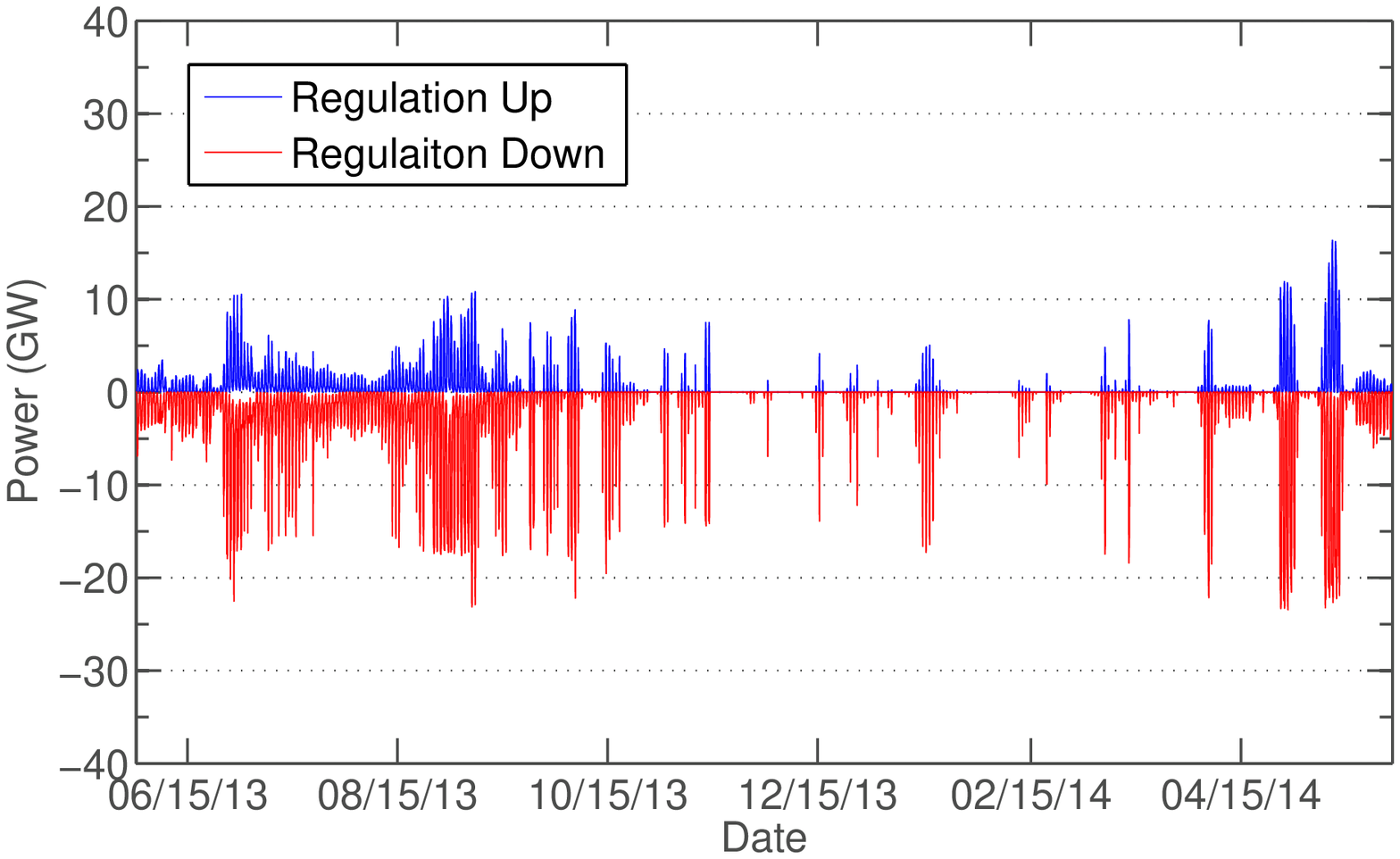}}
\subfigure[Hourly energy capacity of ACs]{\includegraphics[width=.49\columnwidth]{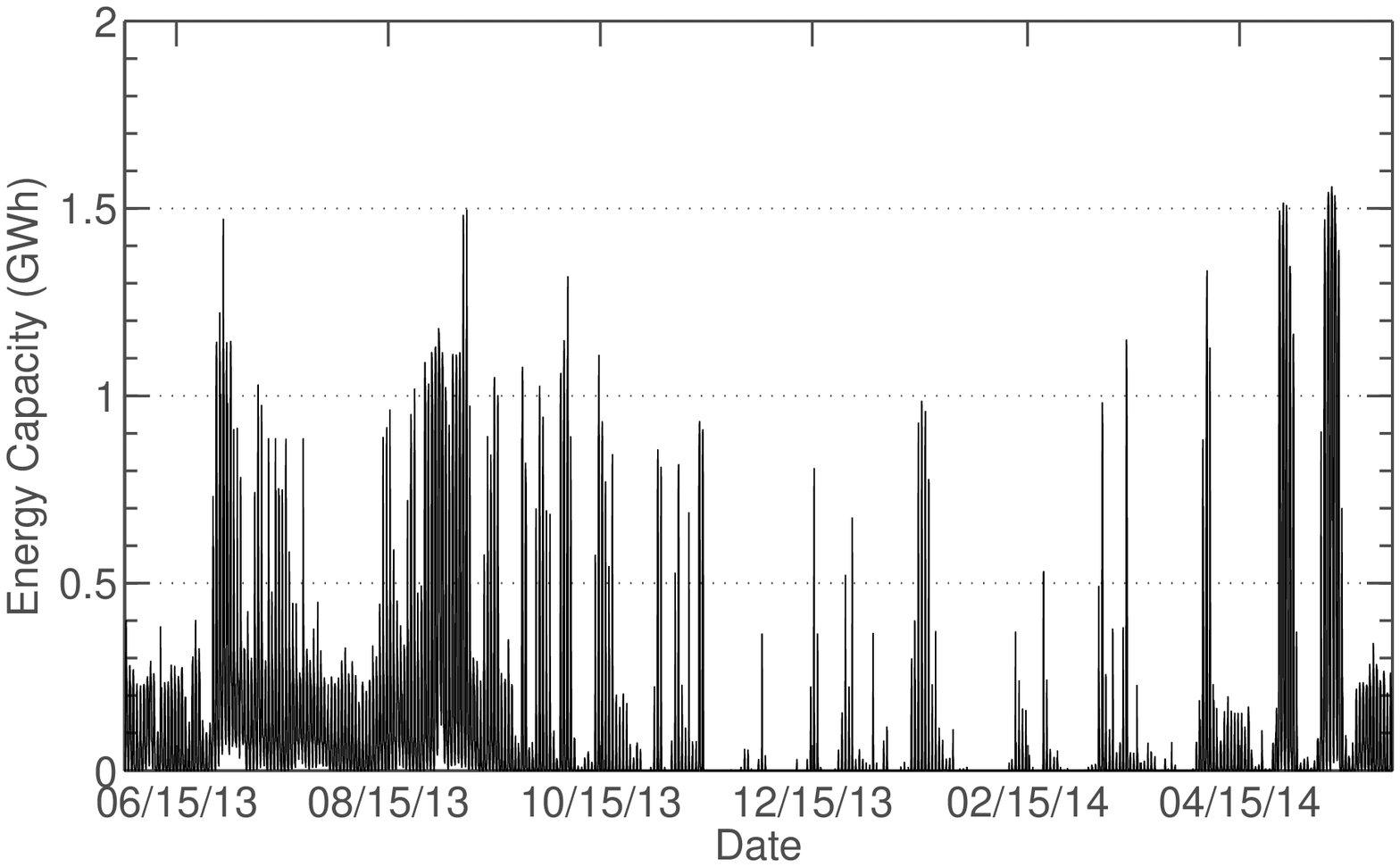}}
\subfigure[Hourly power limits of heat pumps]{\includegraphics[width=.49\columnwidth]{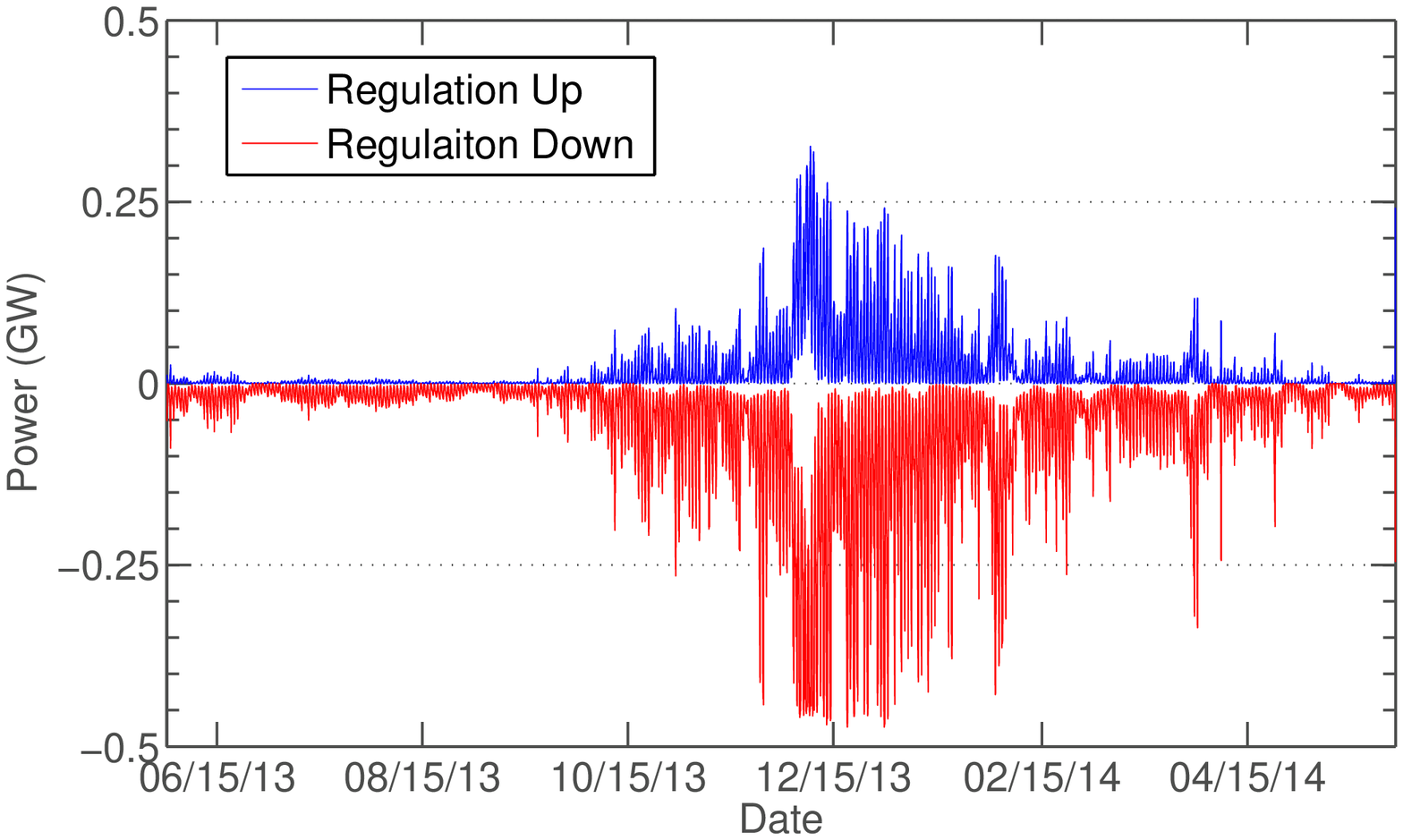}}
\subfigure[Hourly energy capacity of heat pumps]{\includegraphics[width=.49\columnwidth]{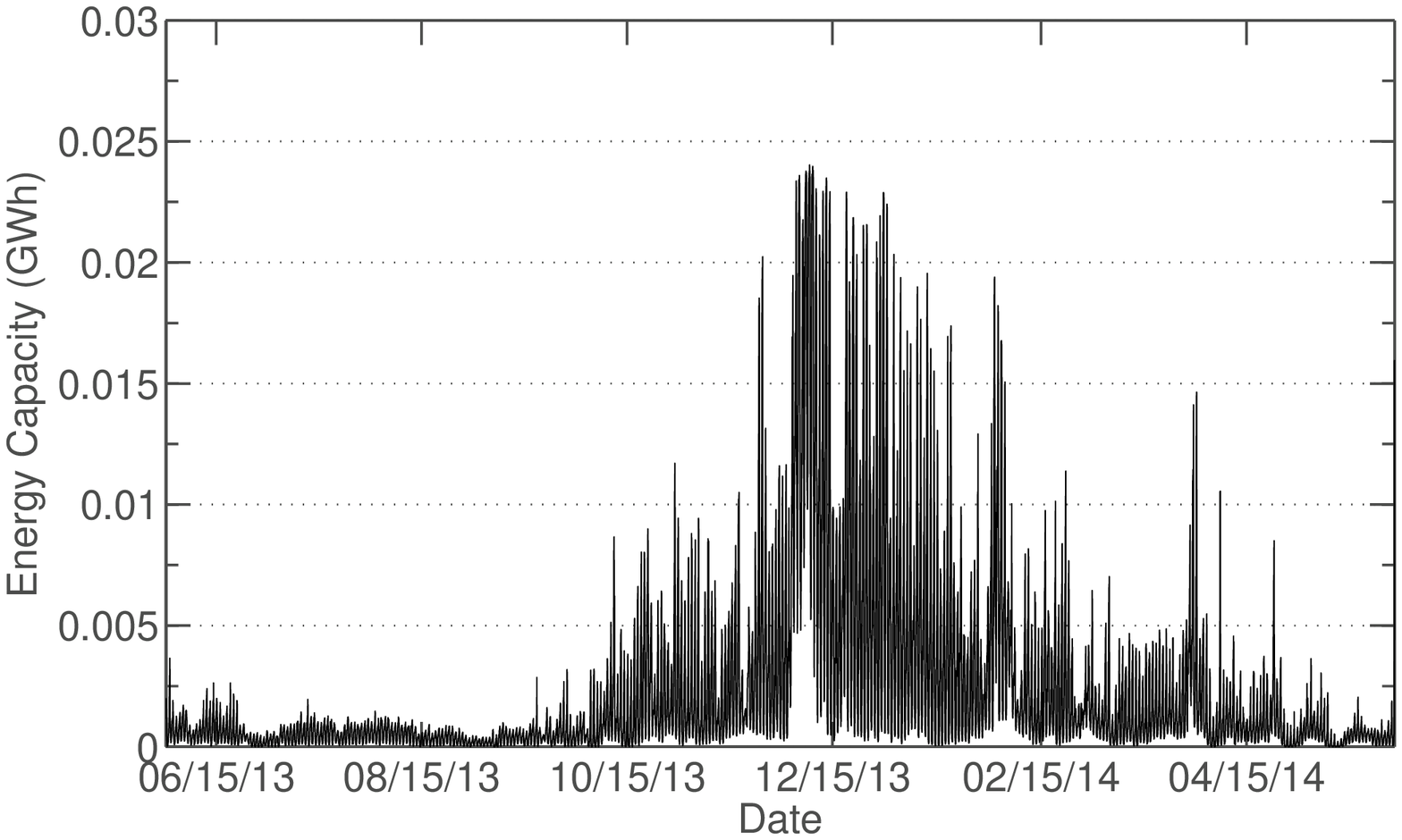}}
\caption{Weighted hourly power limits and energy capacities of ACs (a, b) and heat pumps (c, d) in California from June 2013 to May 2014. }\label{fig:yearly_capacity}
\end{figure}

The weighted hourly power limits and energy capacities of ACs and heat pumps for the 8760 hours from June 2013 to May 2014 are depicted in Fig.  3. Additionally, the average power limits and energy capacities of \acp{TCL} using their annual hourly average temperature profiles in a 24-hour period are depicted in Fig.  4. The colored regions in Fig.  4 represent the total power limits and energy capability from the four types of TCLs. Since the ambient temperature of water heaters and refrigerators are assumed to be constant, their 8760-hour power limits and energy capacities are constants, which are the same as those shown in Fig.  4. 

\cut{
\begin{figure}[tb]
\centering
\subfigure[Regulation up power limit]{\includegraphics[width=.325\columnwidth]{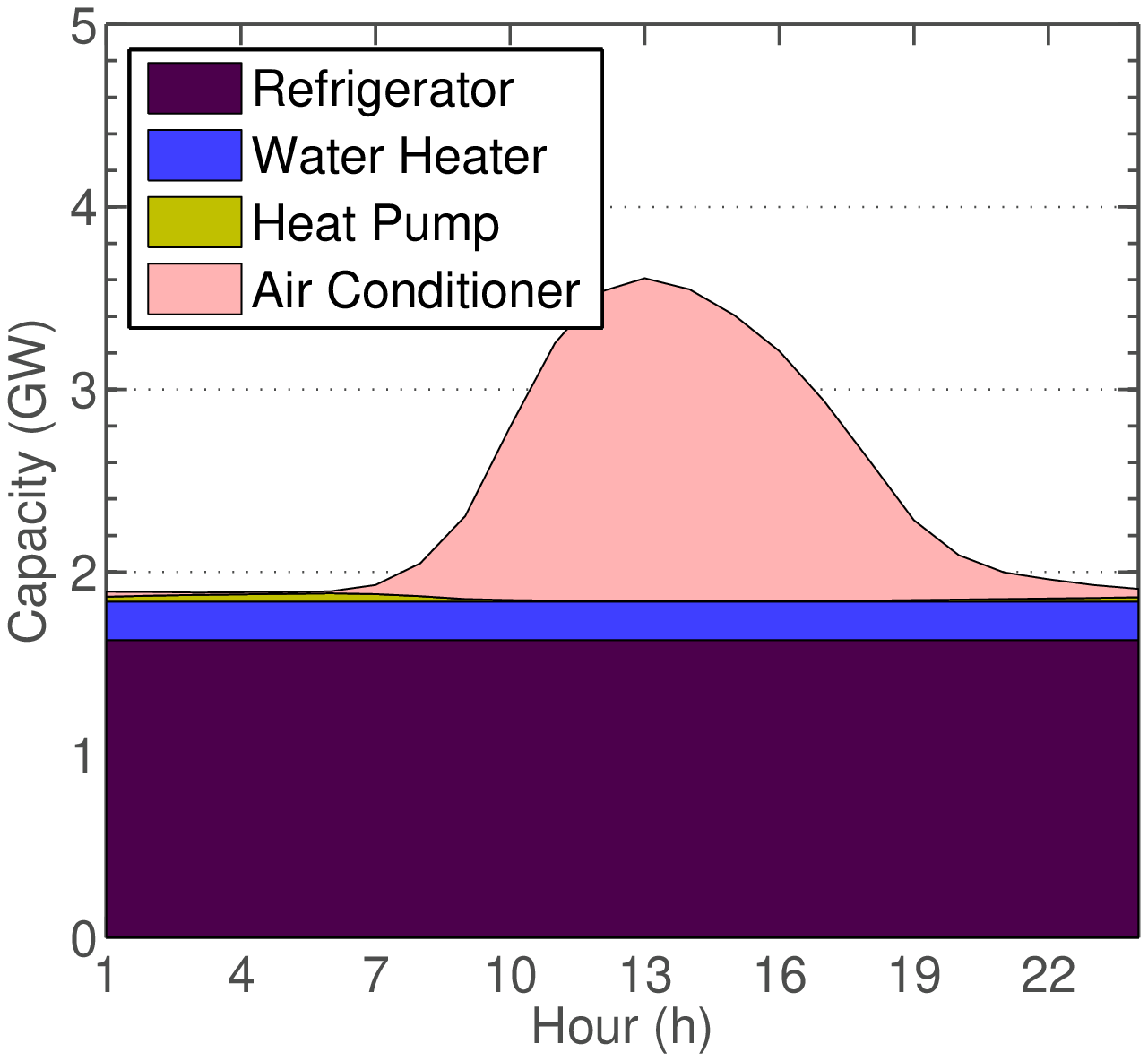}}
\subfigure[Regulation down power limit]{\includegraphics[width=.325\columnwidth]{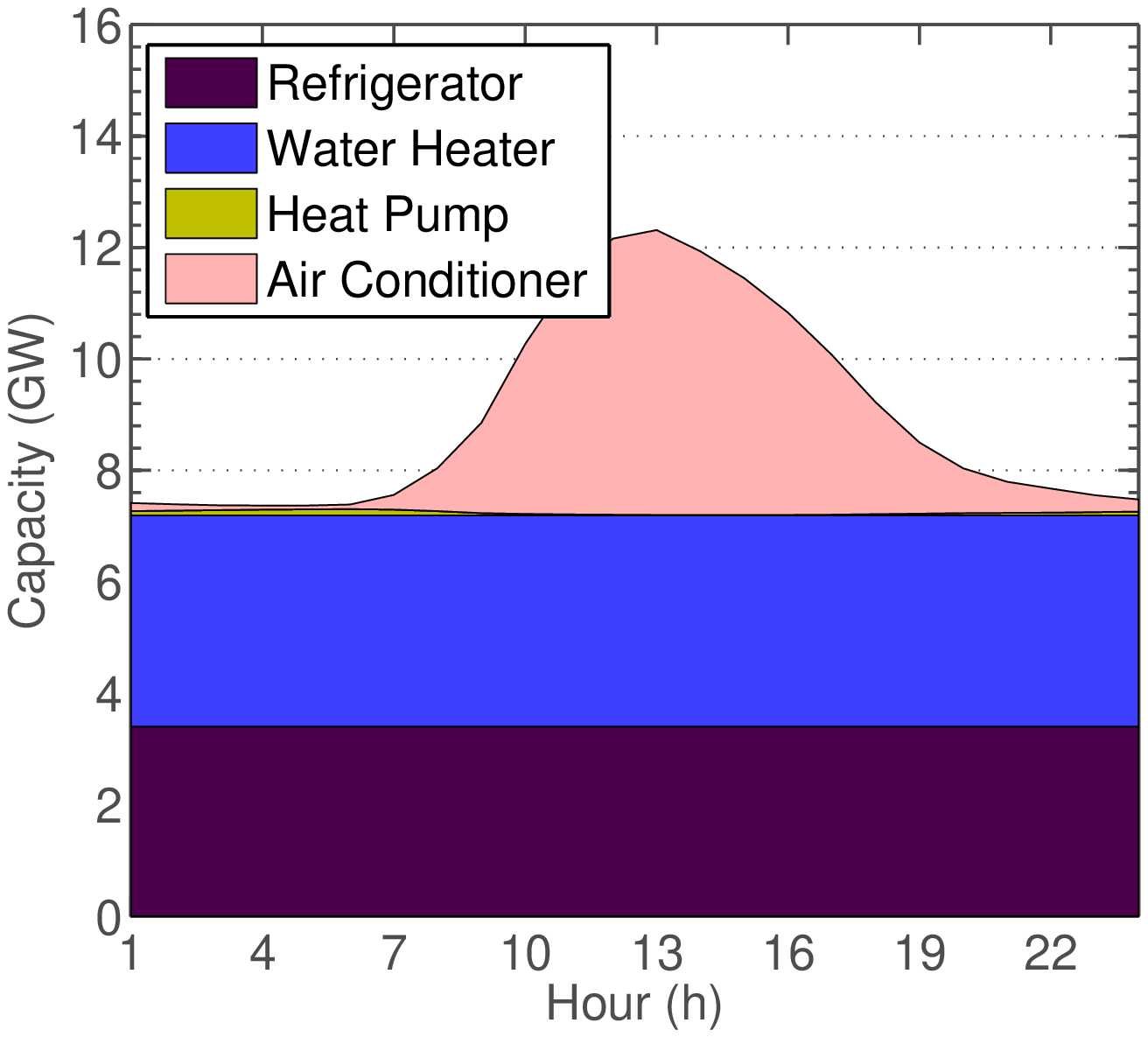}}
\subfigure[Energy Capacity]{\includegraphics[width=.325\columnwidth]{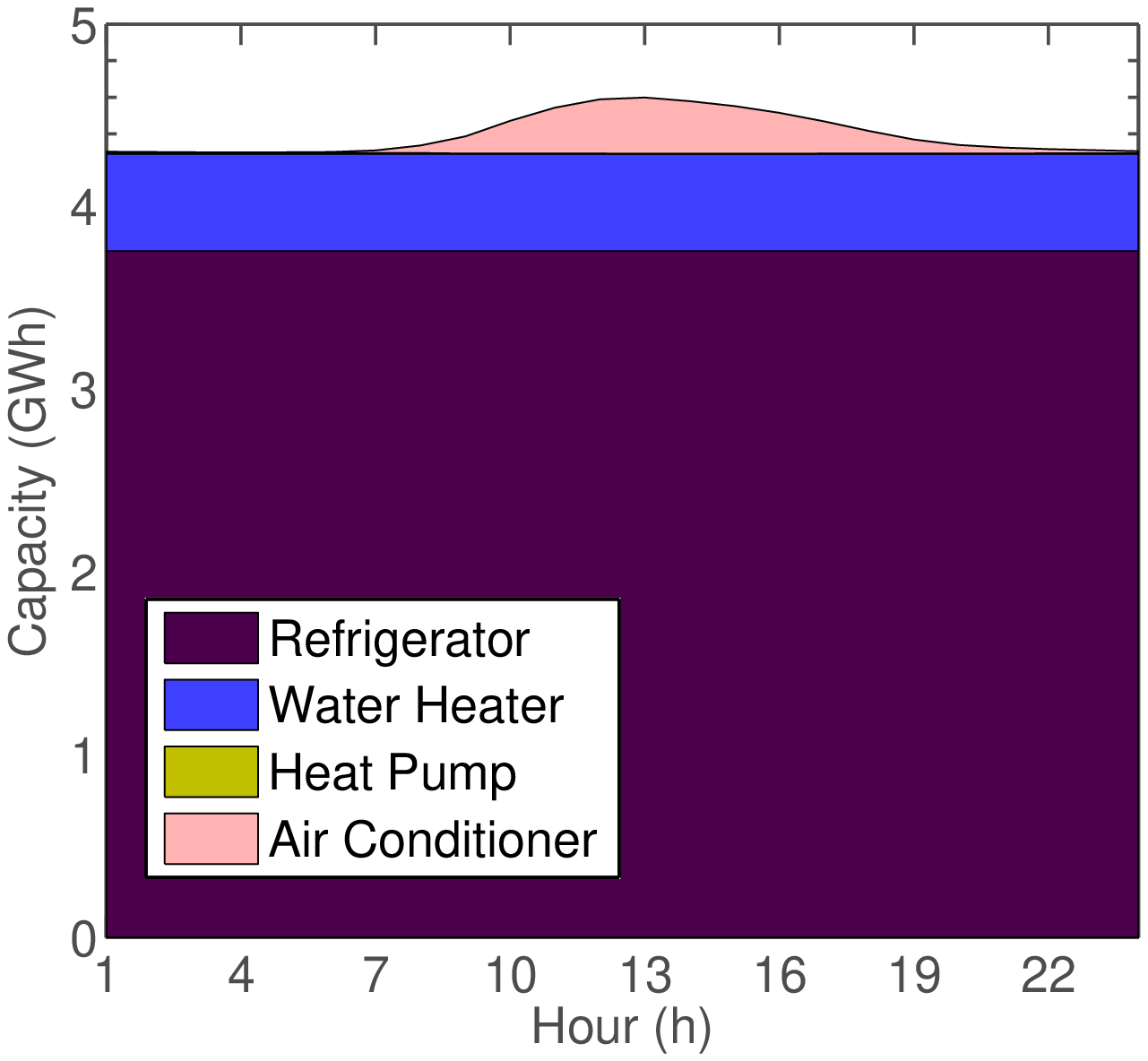}}
\caption{Hourly average upward (a), downward (b) power limits, and energy capacity (c) of TCLs in California  from June 2013 to May 2014. 
}
\label{fig:est_capacity}
\end{figure}
}

We comment that carefully selection of residential customers for demand response is also an important future work. If we assume all TCLs are in Sacramento, which is very hot in summer and cold in winter, ACs and heat pumps present a much larger potential. For example, the power limits and energy capacities of ACs and heat pumps for the 8760-hour period from June 2013 to May 2014 using the temperature profile in Sacramento are depicted in Fig.  5. Compared to Fig.  3, the potential is much larger and more concentrated. 

\begin{figure}[tb]
\centering
\subfigure[Regulation up power limit]{\includegraphics[width=.325\columnwidth]{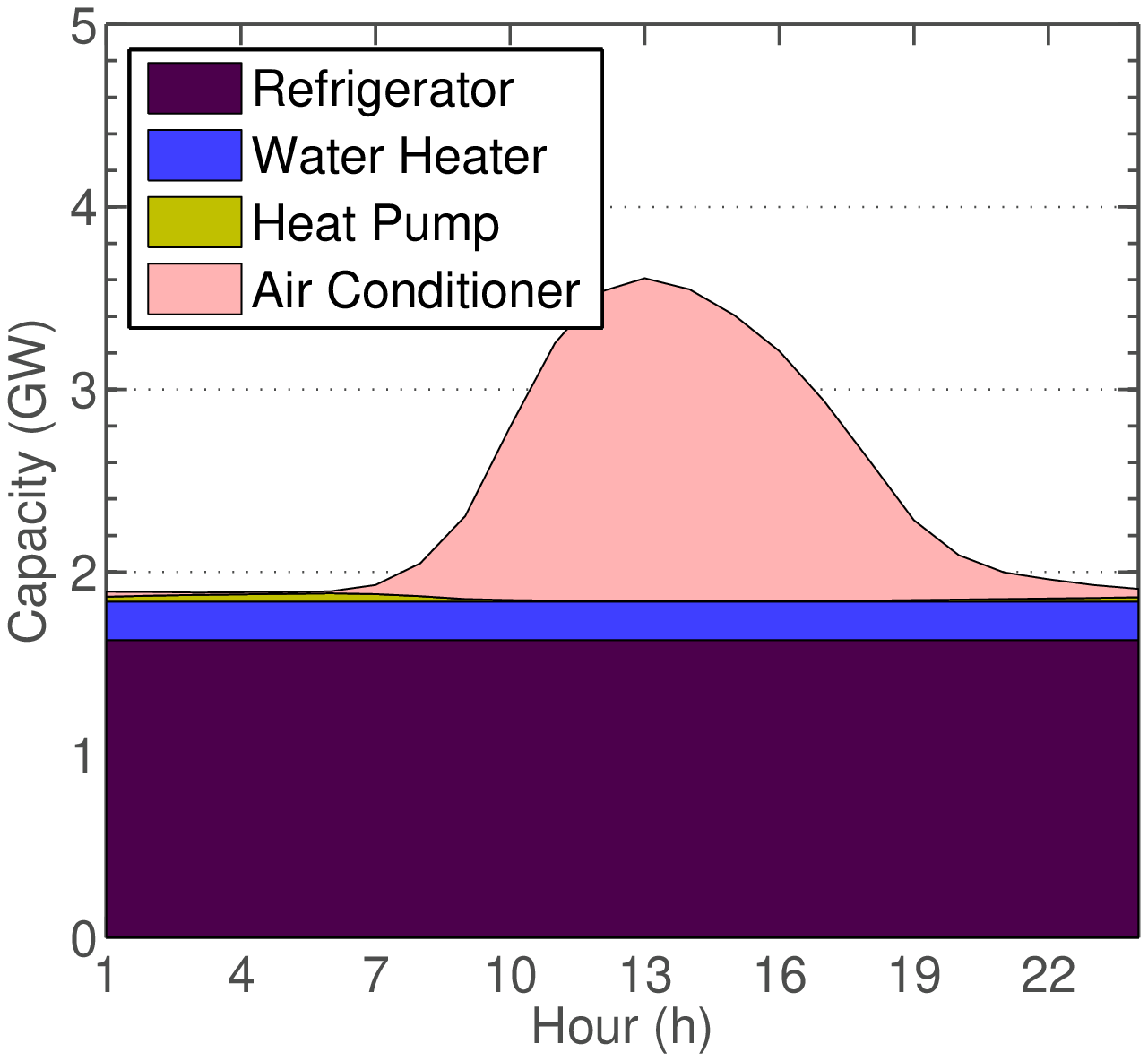}}
\subfigure[Regulation down power limit]{\includegraphics[width=.325\columnwidth]{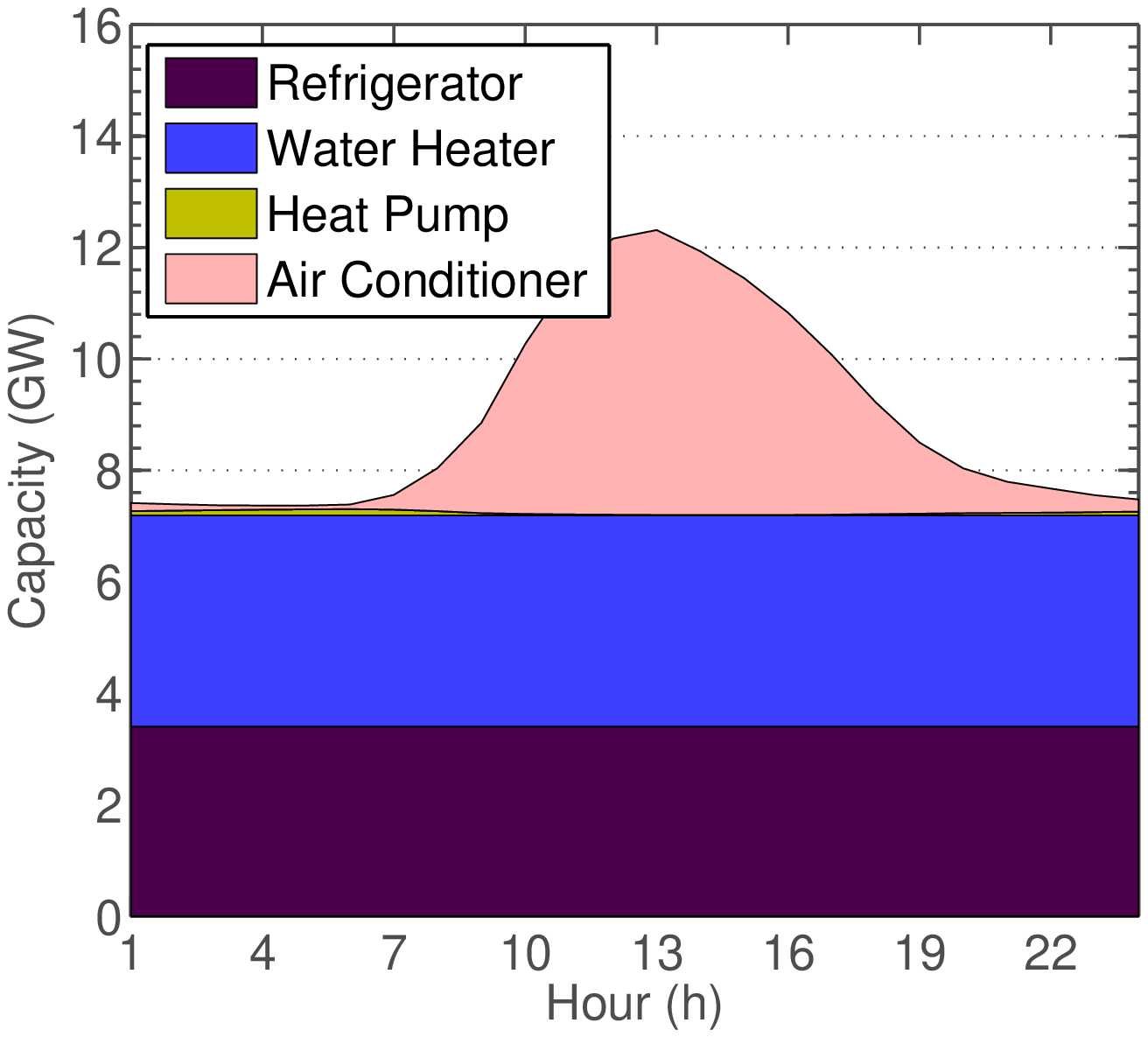}}
\subfigure[Energy Capacity]{\includegraphics[width=.325\columnwidth]{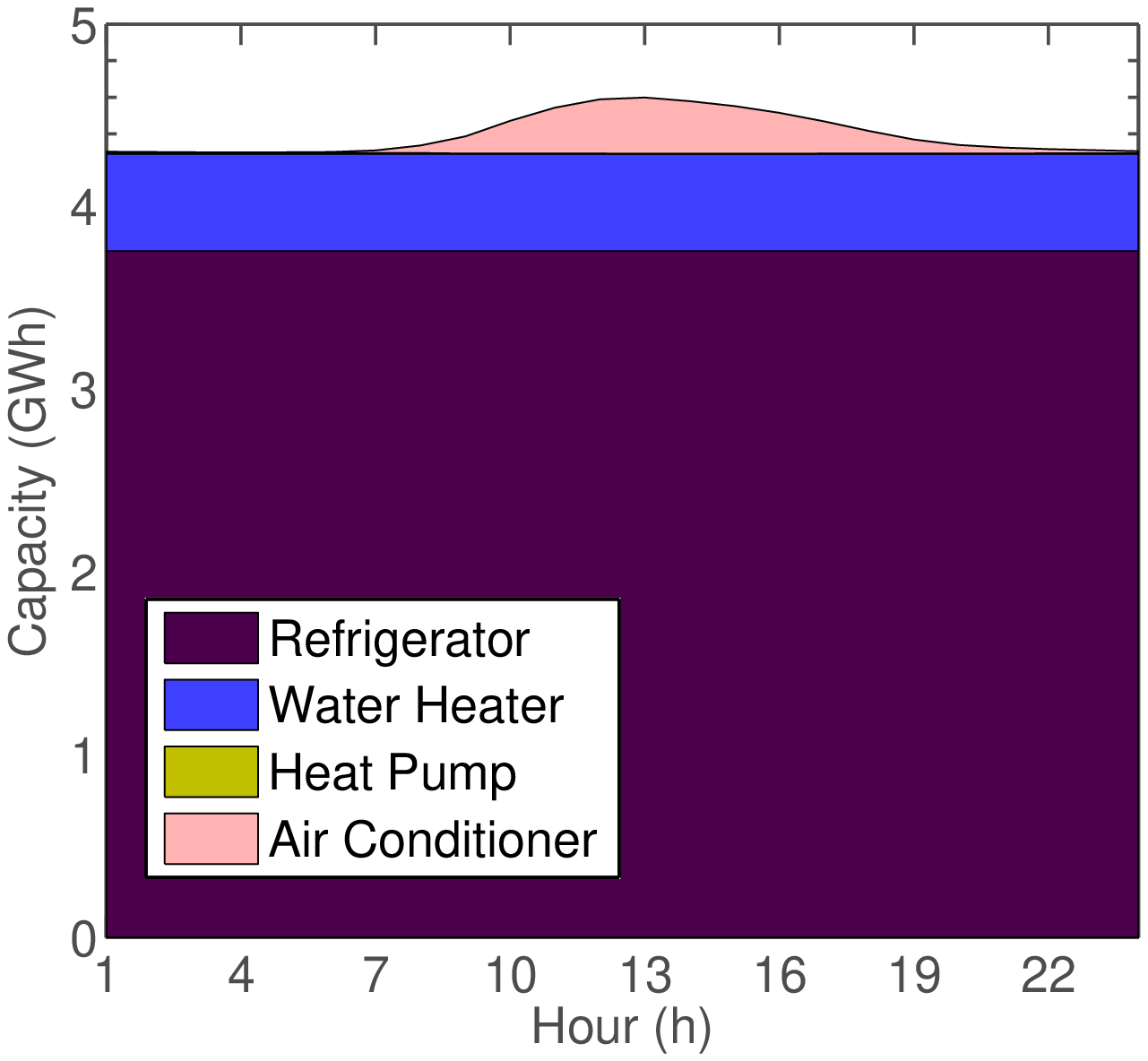}}
\caption{Hourly average upward (a), downward (b) power limits, and energy capacity (c) of TCLs in California  from June 2013 to May 2014. 
}
\label{fig:est_capacity}
\end{figure}

In Table III, the peak values for the power limits and energy capacities (such as those shown in Fig.  4) are given using the annual hourly average temperature profiles for the five different cities. The final row shows the minimum values of the total power limits and energy capacity over the 24-hour period. We observe from Table III that even using the most conservative temperature profile (SF), the minimum power limits are more than twice of the current maximum regulation procurement ($600$ MW). Furthermore, it is larger than the  predicted regulation procurement ($1.3$ GW) if California achieved its $33\%$ of renewable penetration by 2020. 

\begin{figure}[tb]
\centering
\subfigure[Hourly power limits of ACs]{\includegraphics[width=.49\columnwidth]{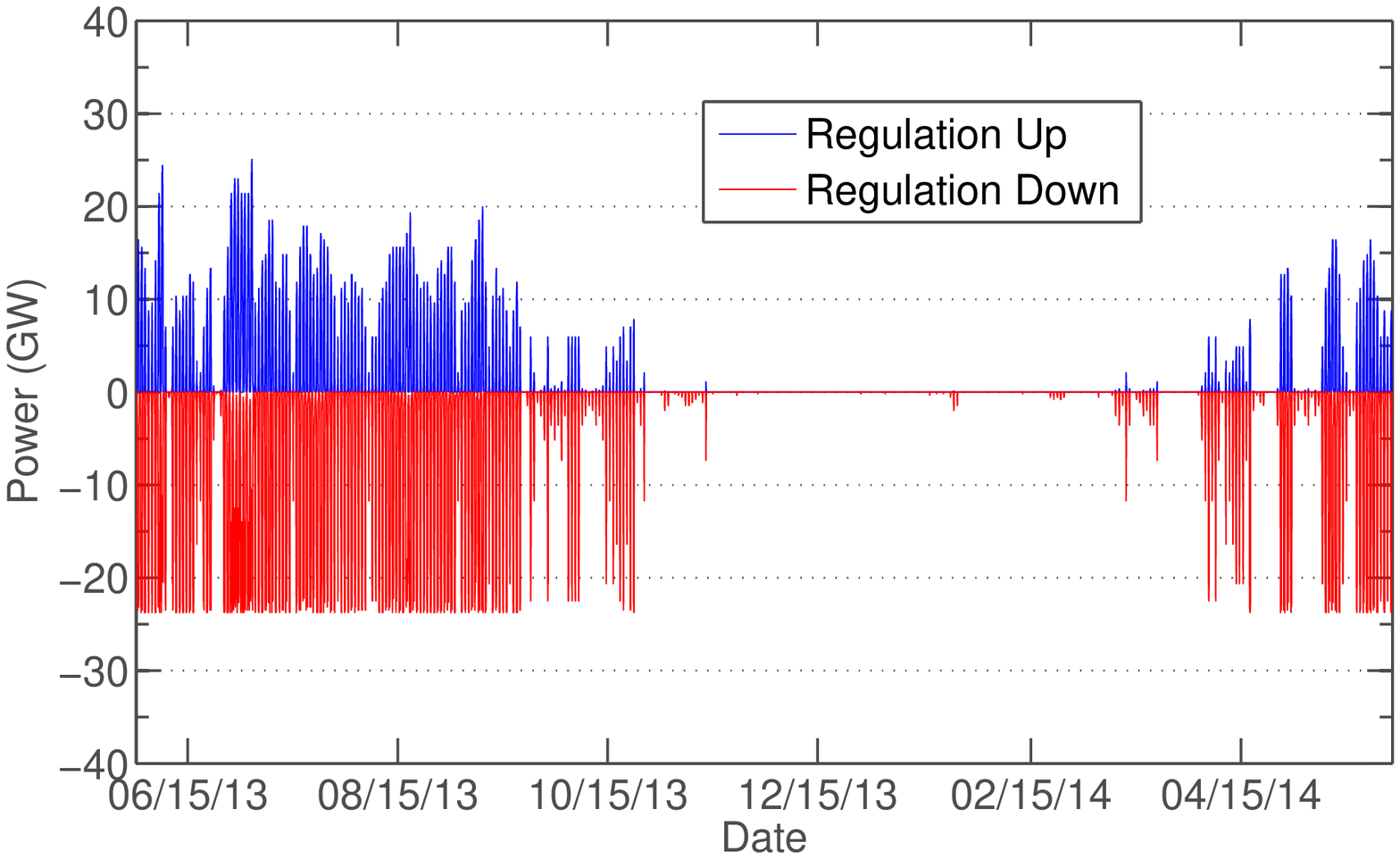}}
\subfigure[Hourly energy capacity of ACs]{\includegraphics[width=.49\columnwidth]{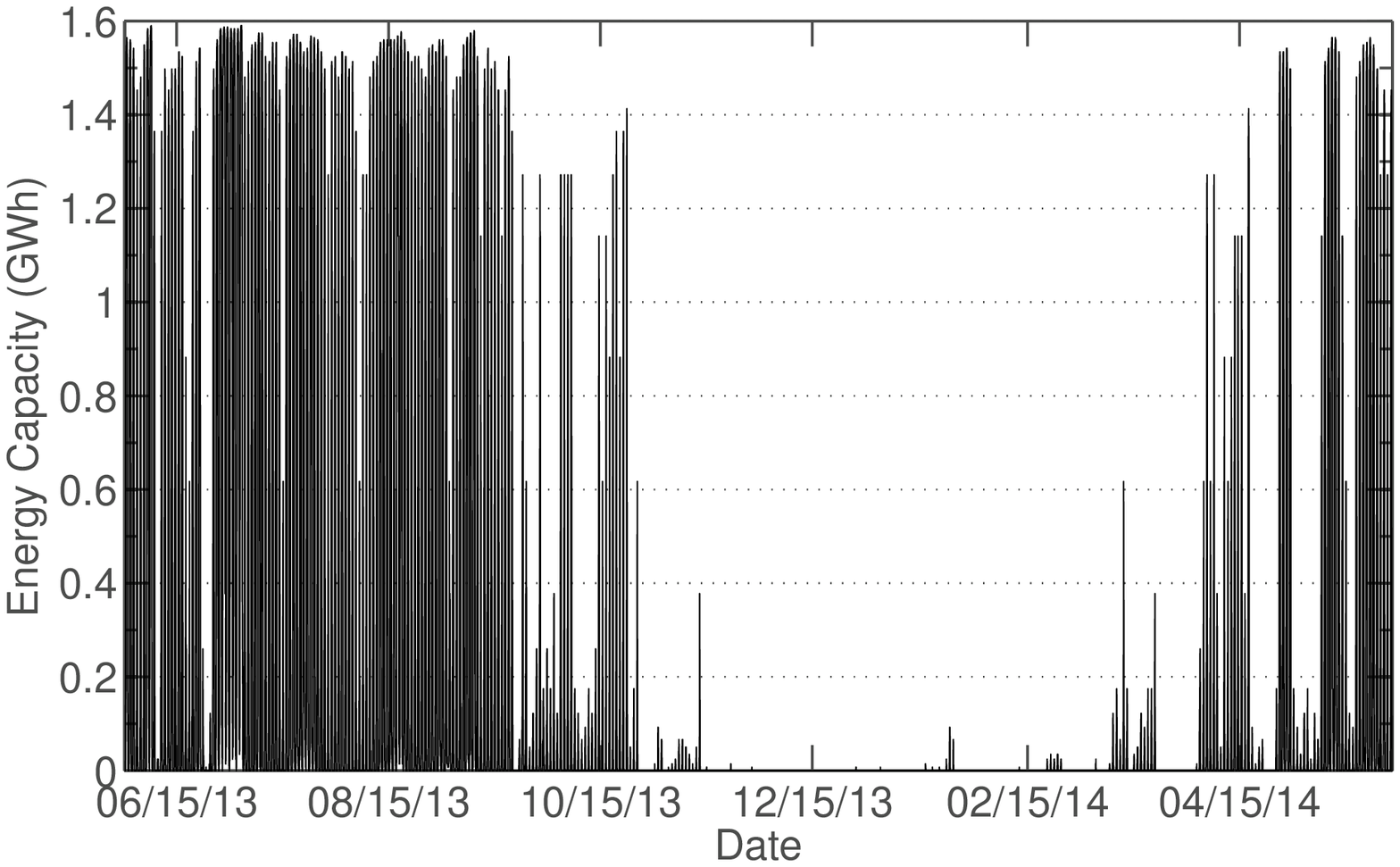}}
\subfigure[Hourly power limits of heat pumps]{\includegraphics[width=.49\columnwidth]{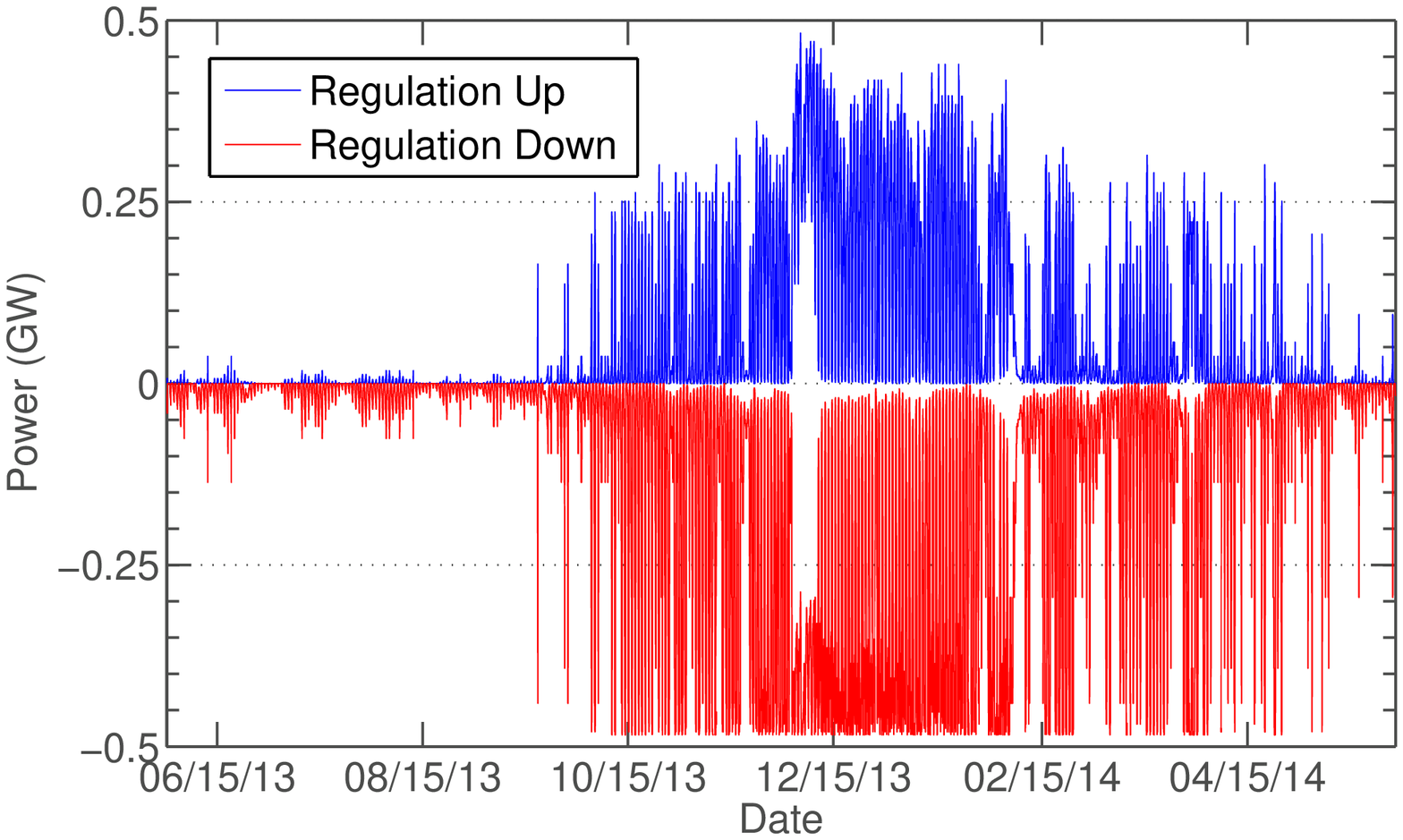}}
\subfigure[Hourly energy capacity of heat pumps]{\includegraphics[width=.49\columnwidth]{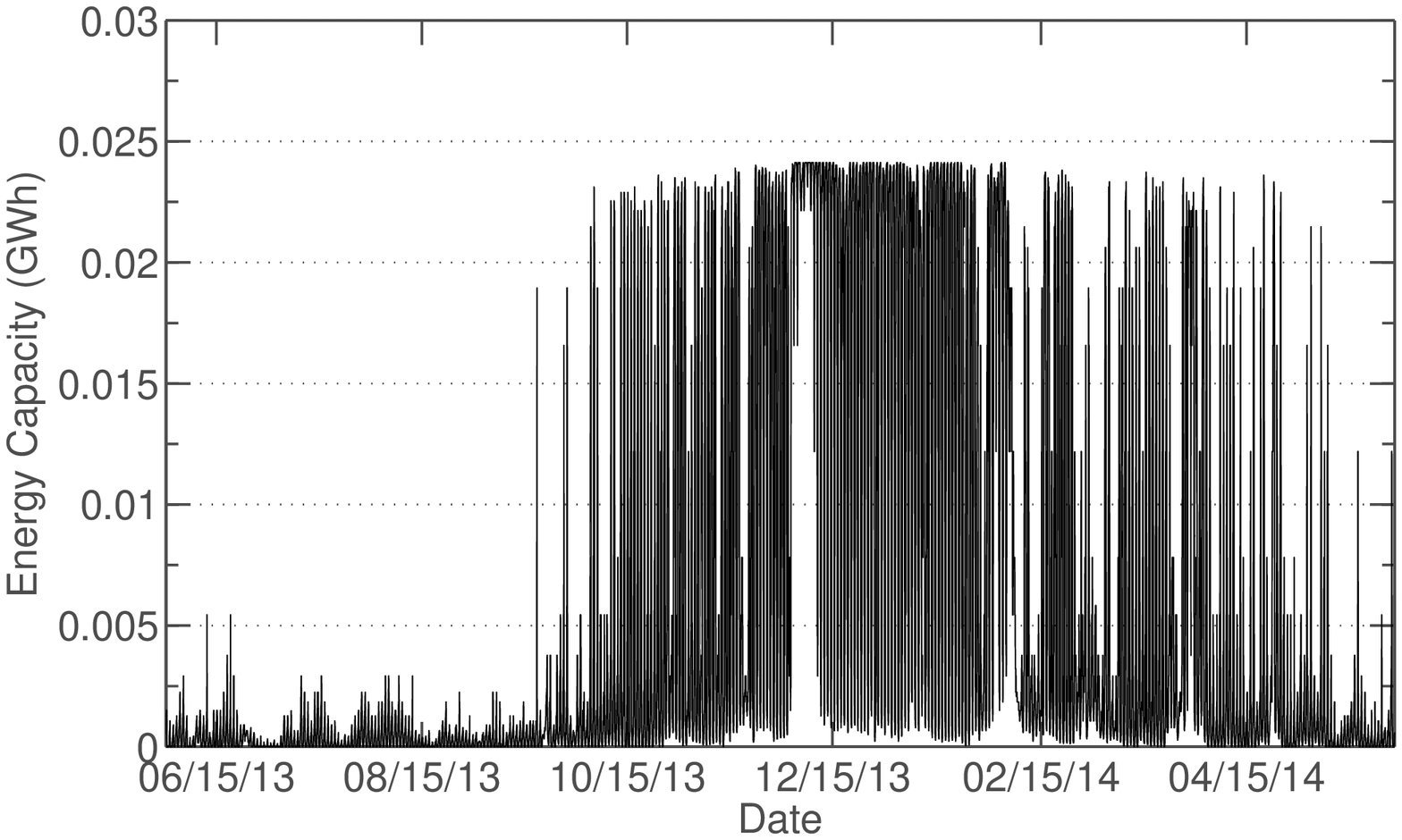}}
\caption{Hourly power limits and energy capacities of ACs (a, b) and heat pumps (c, d) using the annual temperature profile in Sacramento from June 2013 to May 2014. }\label{fig:yearly_capacity_SA}
\end{figure}

Additionally, assuming the regulation signal in \ac{CAISO} has similar pattern as that in the \ac{PJM} Interconnection, we use the regulation signal of \ac{PJM} (which is available in \cite{PJM}) to estimate the maximum energy requirement for regulation provision using Definition \ref{def:battery}. We assume for each type of \ac{TCL}, the dissipation rate $\alpha$ of the battery model associated with them is equal to the average of their model parameter $a$'s. From Table I, we see that  the average time constants of ACs, heat pumps, water heaters, and refrigerators are respectively $0.25$ h$^{-1}$, $0.25$ h$^{-1}$, $0.02$ h$^{-1}$, and $0.02$ h$^{-1}$. We first estimate the energy requirements for ACs and heat pumps. Using the battery model (given in Definition \ref{def:battery}) with maximum power, minimum power, and dissipation parameters fixed to be $600$ MW, $600$ MW, and $0.25$ h$^{-1}$ respectively, we estimate the maximum energy requirement $\mathcal{C}^{\rm max}$ by integrating the associated battery model $ \dot{x}(t) = -0.25 x(t) -  600r(t)$  from $t=0$ to $t=24$, where $r(t)$ is a 24-hour long normalized regulation signal from PJM, and finding the maximum absolute value of $x(t)$ over the 24-hour period. More precisely, we numerically compute,
\begin{align*}
\mathcal{C}^{\rm max}  =  \max |x(t)|,  \quad \text{where} \ \dot{x}(t) = -0.25 x(t) - 600 r(t), \ x(0)=0, \ \forall \ t \in [0,\ 24]. 
\end{align*}
The estimated maximum energy requirement for regulation with the $600$ MW (up and down) power procurement is about $150$ MW-h. Additionally, the estimated maximum energy requirement for regulation with a $1.3$ GW (up and down) power procurement is about $350$ MW-h. Similarly, we estimate the maximum energy requirements for water heaters and refrigerators. The maximum energy requirements with $600$ MW and $1.3$ GW power procurements are respectively $180$ MW-h and $380$ MW-h. We observe from the last row of Table III that the energy requirements are much smaller than the available energy capacity from \acp{TCL}.

Altogether, these results show that the potential of \acp{TCL} in California is more than enough for provision of regulation reserve for now and the near future. 

\begin{table}[tb]
\caption{Potential of \acp{TCL} using different temperature profiles in Sacramento (SA), San Francisco (SF), Bakersfield (BF), Los Angeles (LA), and San Diego (SD).}
\label{tab:potential}
\vspace{6 pt}
\centering
{\scriptsize
\begin{tabular}{c c|| c| c | c | c | c}
 & & SA & SF & BF & LA & SD \\
 \hline \hline
\multirow{3}{*}{AC (peak)} & Regulation up (GW) &  4.65 & 0.22 & 6.45 & 1.96 &0.42 \\
 & Regulation down (GW)  &  9.12 & 0.90 & 10.2& 6.14 & 1.38 \\
 & Energy Capacity (GWh)  &  0.61 & 0.05 & 0.75 & 0.36 & 0.08 \\
 & Dissipation rate (h$^{-1}$) & 0.25 & 0.25 & 0.25 & 0.25 & 0.25 \\
\hline
\multirow{3}{*}{Heat Pump (Peak)} & Regulation up (GW)  &  0.13 & 0.06 & 0.08 & 0.03 & 0.04 \\
 & Regulation down (GW)  &  0.23 & 0.16 & 0.17 & 0.09 & 0.11 \\
 & Energy Capacity (GWh)  &  .011 & .007 & .007 & .004 & .005 \\
 & Dissipation rate (h$^{-1}$) & 0.25 & 0.25 & 0.25 & 0.25 & 0.25 \\
\hline
\multirow{3}{*}{Water Heater} & Regulation Up (GW)  &  0.21 & 0.21 & 0.21 & 0.21 &0.21\\
 & Regulation down (GW) &  3.79 & 3.79 & 3.79 & 3.79 & 3.79 \\
 & Energy Capacity (GWh) &  0.53 & 0.53 & 0.53 & 0.53 &0.53 \\
  & Dissipation rate (h$^{-1}$) & 0.02 & 0.02 & 0.02 & 0.02 & 0.02 \\
\hline
\multirow{3}{*}{Refrigerator} & Regulation up  (GW) &  1.63 & 1.63 & 1.63 & 1.63 &1.63 \\
 & Regulation down (GW)  &  3.38 & 3.38 & 3.38 & 3.38 &3.38 \\
 & Energy Capacity (GWh) &  3.78 & 3.78 & 3.78 & 3.78 &3.78 \\
  & Dissipation rate (h$^{-1}$) & 0.02 & 0.02 & 0.02 & 0.02 & 0.02 \\
\hline \hline
 \multirow{3}{*}{\textbf{Total (minimum)}} & Regulation up  (GW) &  \textbf{1.94} & \textbf{1.86} & \textbf{2.08} & \textbf{1.86} & \textbf{1.86} \\
 & Regulation down (GW)  &  \textbf{7.42} & \textbf{7.28} & \textbf{8.20} & \textbf{7.30} & \textbf{7.28} \\
 & Energy Capacity (GWh) &  \textbf{4.31} & \textbf{4.31} & \textbf{4.34} & \textbf{4.30} & \textbf{4.31}
\end{tabular}}
\end{table}

\subsection{Revenue Analysis of \acp{TCL} Providing Regulation Reserve}
We next estimate the potential revenue of \acp{TCL} for frequency regulation using historic data of CAISO. In the ancillary service market, frequency regulation is the most expensive service. The market clearing prices of other ancillary services such as contingency reserves (spinning or non-spinning) and supplemental reserve are much cheaper, compared with frequency regulation \cite{AS_Kirby}.

\begin{figure}[tb]
\centering
\subfigure[Regulation Up]{\includegraphics[width=.493\columnwidth]{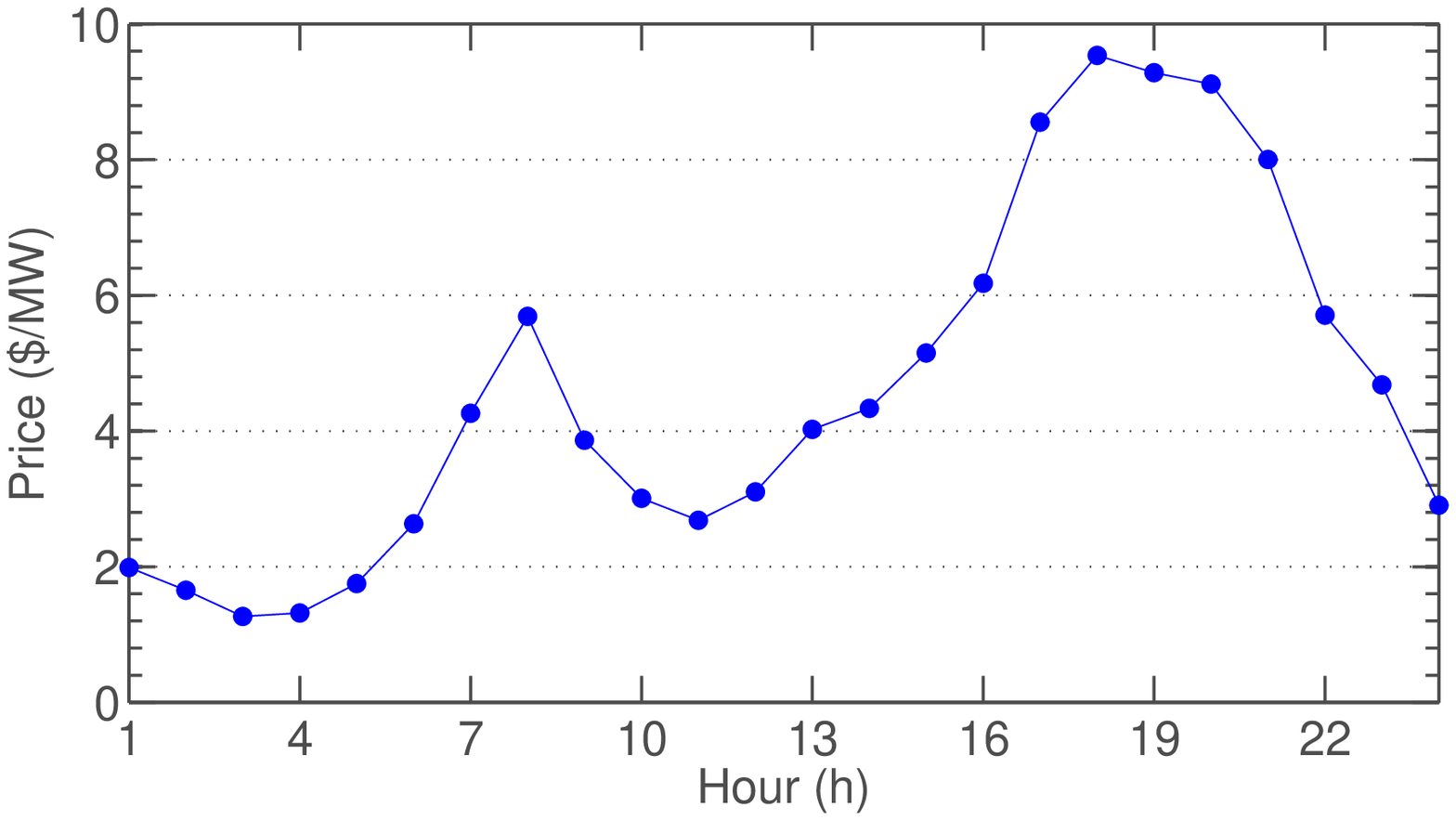}}
\subfigure[Regulation Down]{\includegraphics[width=.493\columnwidth]{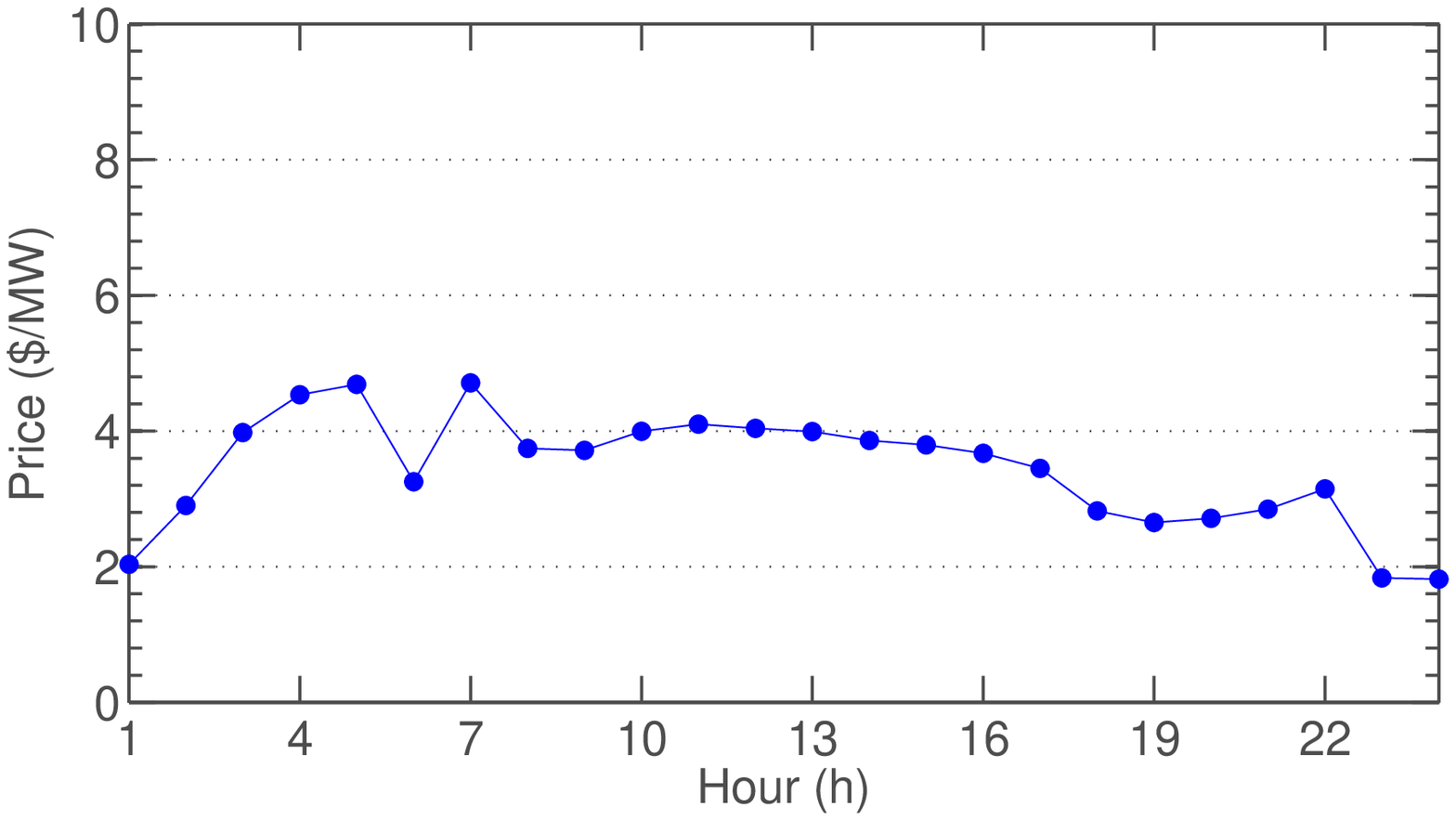}}
\subfigure[Regulation Mileage Up]{\includegraphics[width=.493\columnwidth]{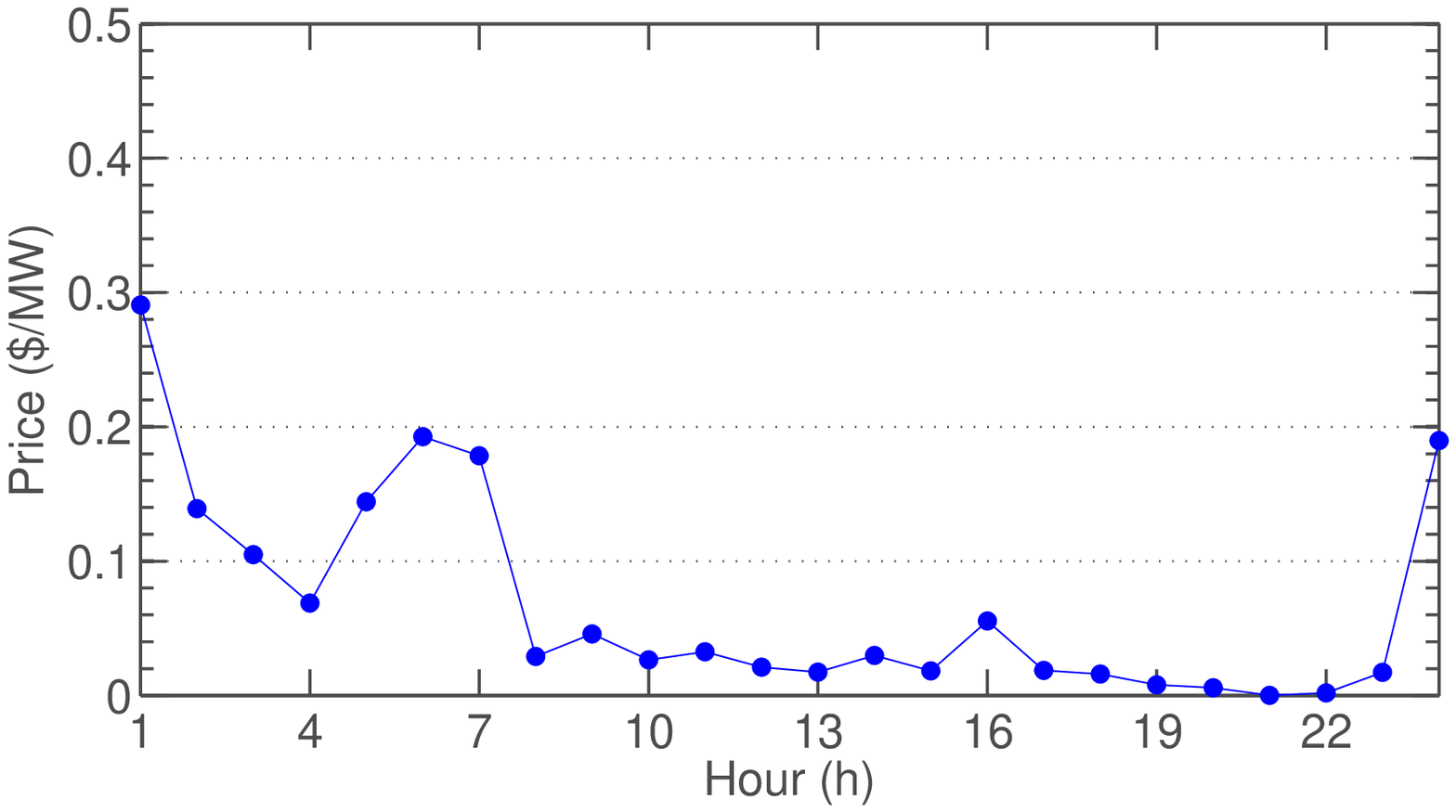}}
\subfigure[Regulation Mileage Down]{\includegraphics[width=.493\columnwidth]{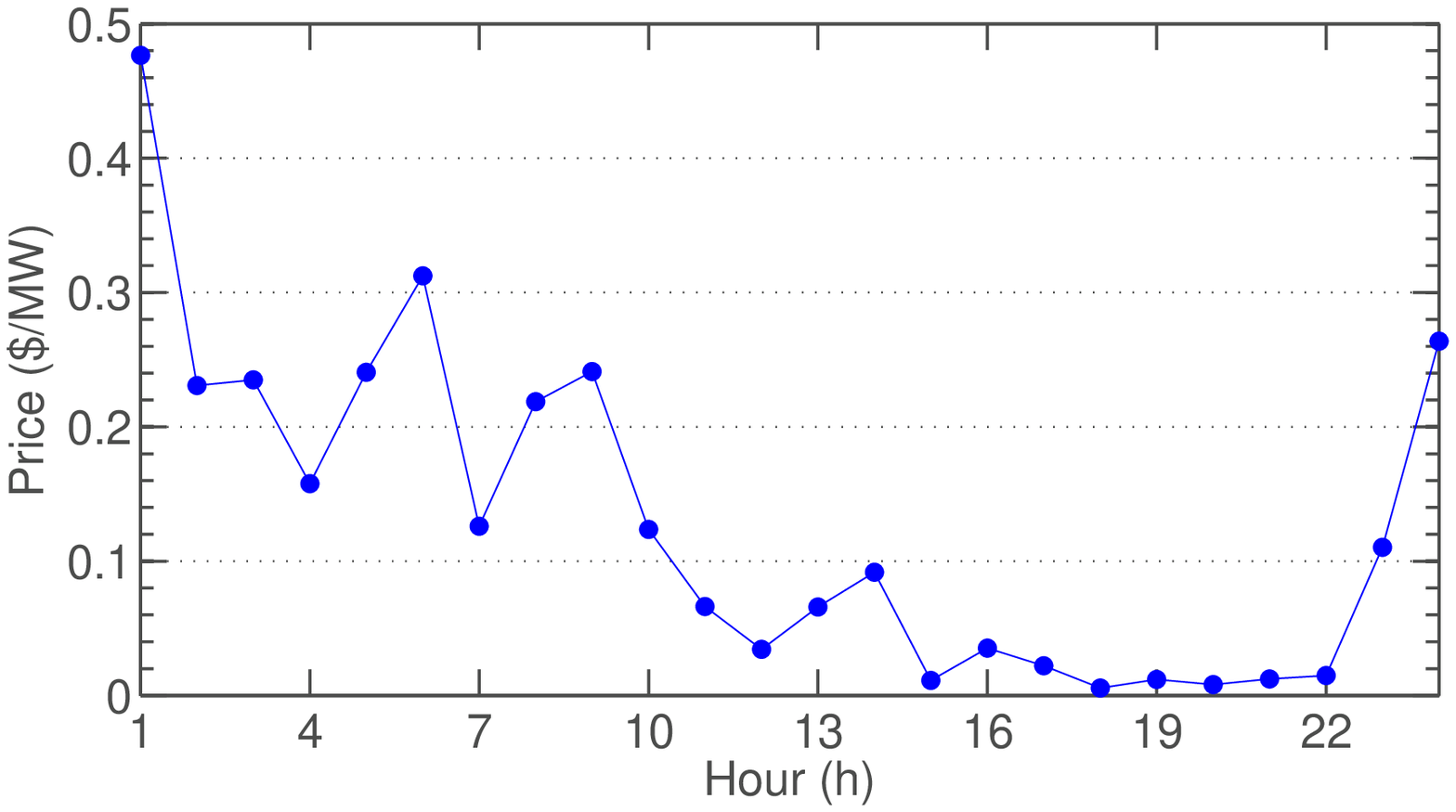}}
\caption{Hourly average MCP for upward and downward regulation in California. The plots are based on historic data of CAISO from June 2013 to May 2014.}\label{fig:price}
\end{figure}

The system operator in \ac{CAISO} clears the regulation service market as follows. First, participating resources submit their offers. The system operator uses these offers together with the energy offers to  determine the lowest cost alternative for these services by conducting a co-optimization. Within the co-optimization, an ISO dispatch profile is created along with Locational Marginal Pricings (LMPs). Using the dispatch profiles and forecast LMPs, an opportunity cost  is estimated for each resource that is eligible to provide regulation. The \ac{MCP} for that operating hour is the sum of the availability bid and opportunity cost associated with the most expensive resource awarded. All awarded resources in a reserve zone are paid the same \ac{MCP}, regardless of their own bid and opportunity cost. Fig.  6 depicts the hourly average MCPs for regulation up, regulation down, regulation mileage up, and regulation mileage down in California from June 2013 to May 2014 \cite{OASIS}. In particular, the average MCPs for regulation up and regulation down in this 8760-hour period were respectively $\$4.61$ and $\$3.43$ per MW, and the average MCP for regulation mileage up and regulation mileage down were respectively $\$0.069$ and $\$0.130$ per MW. 

\begin{figure}[tb]
\centering
\subfigure[Hourly revenue of ACs]{\includegraphics[width=.49\columnwidth]{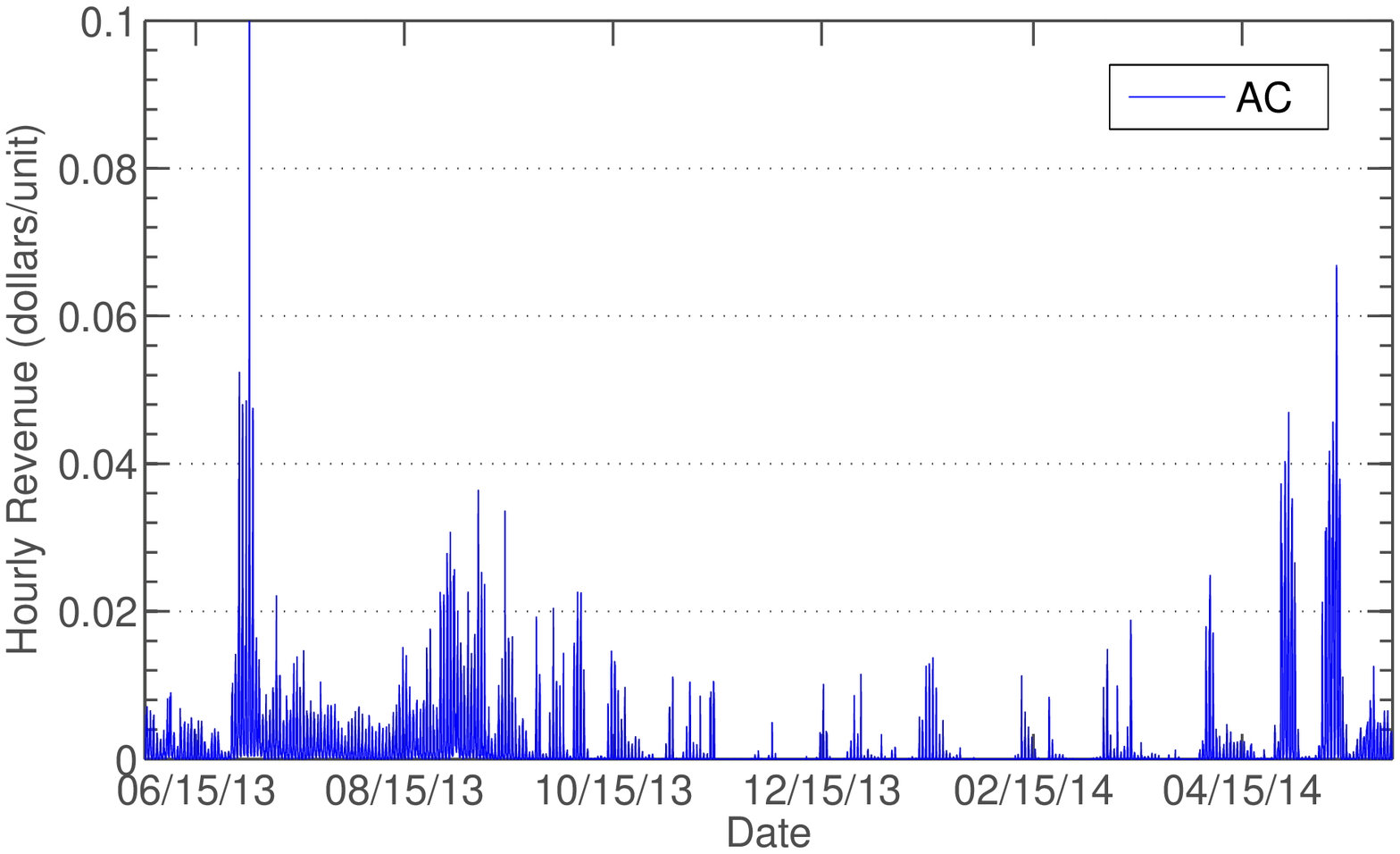}}
\subfigure[Hourly revenue of heat pumps]{\includegraphics[width=.49\columnwidth]{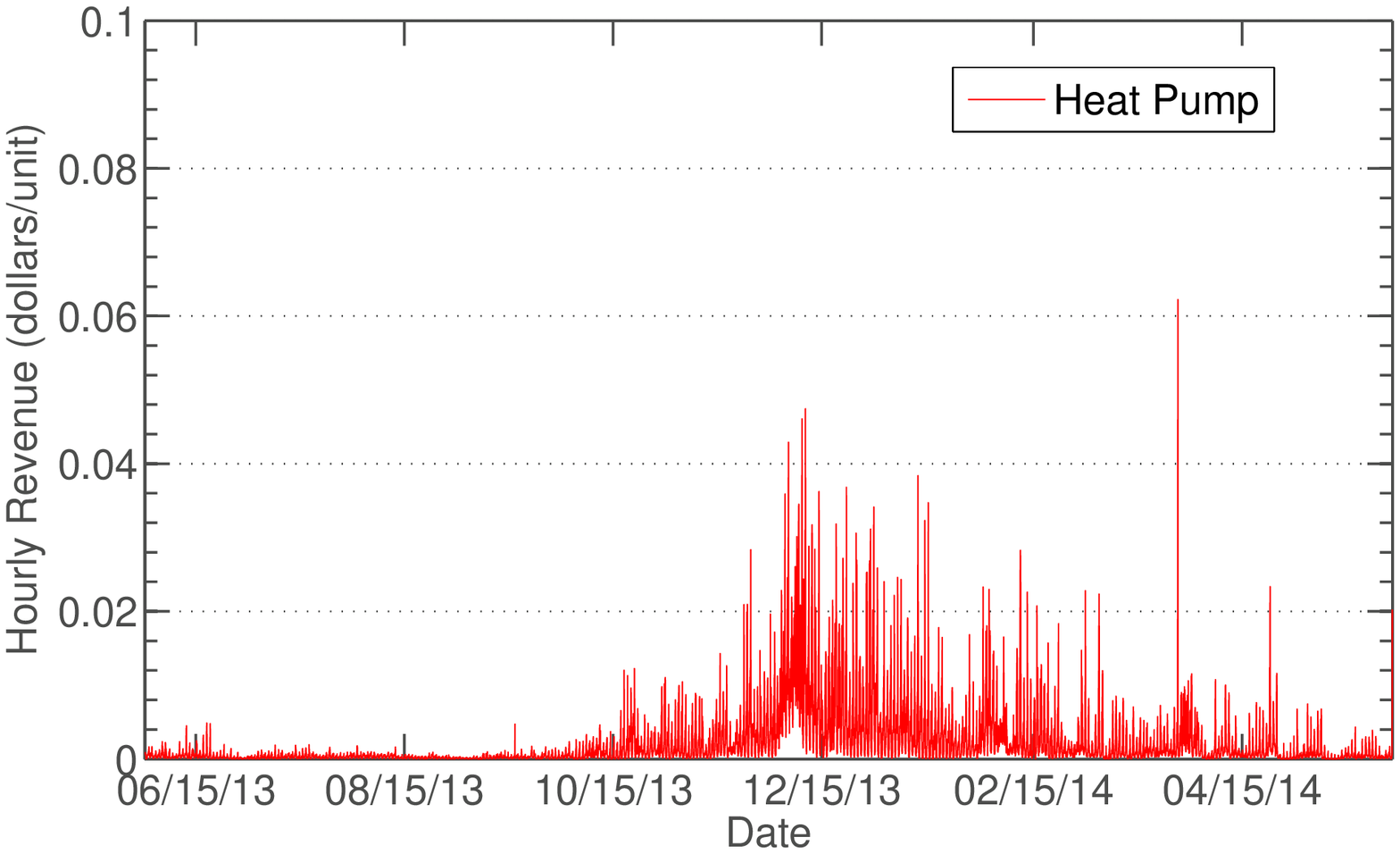}}
\subfigure[Hourly revenue of water heaters]{\includegraphics[width=.49\columnwidth]{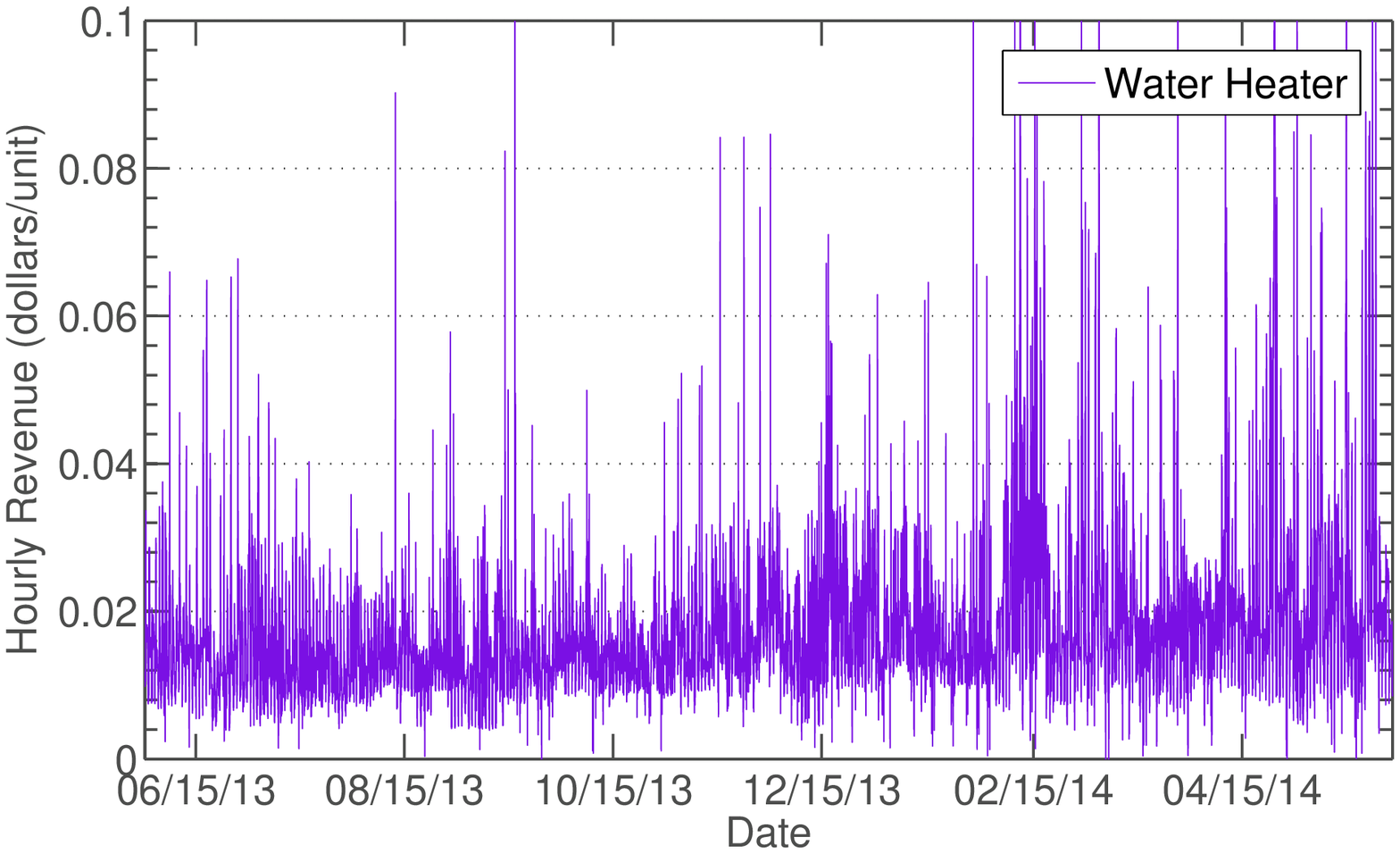}}
\subfigure[Hourly revenue of refrigerators]{\includegraphics[width=.49\columnwidth]{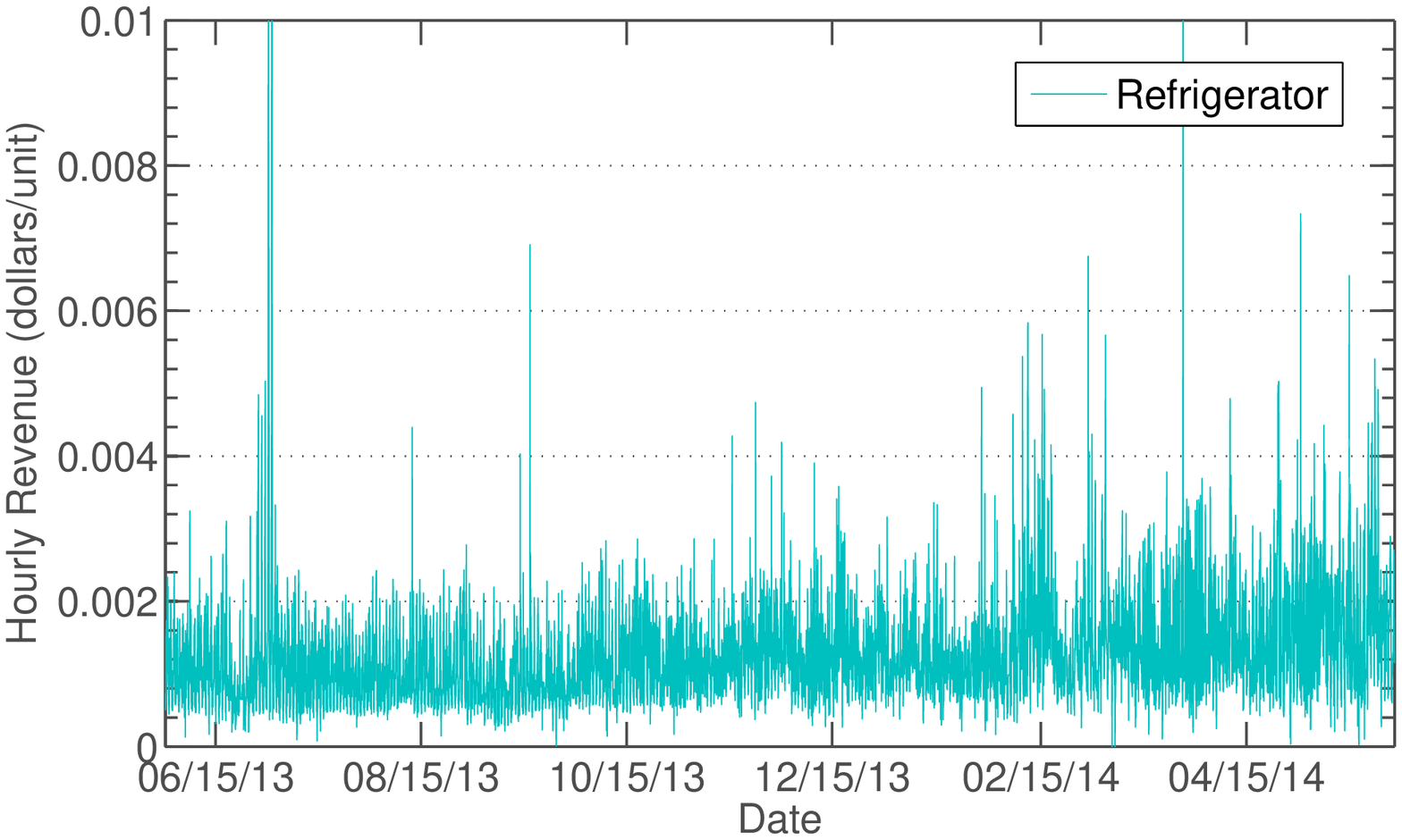}}
\caption{Weighted hourly revenue of ACs (a), heat pumps (b), water heaters (c) and refrigerators (d) in California from June 2013 to May 2014. }\label{fig:yearly_revenue}
\end{figure}

\begin{figure}[tb]
\centering
\subfigure[Hourly revenue of ACs]{\includegraphics[width=.49\columnwidth]{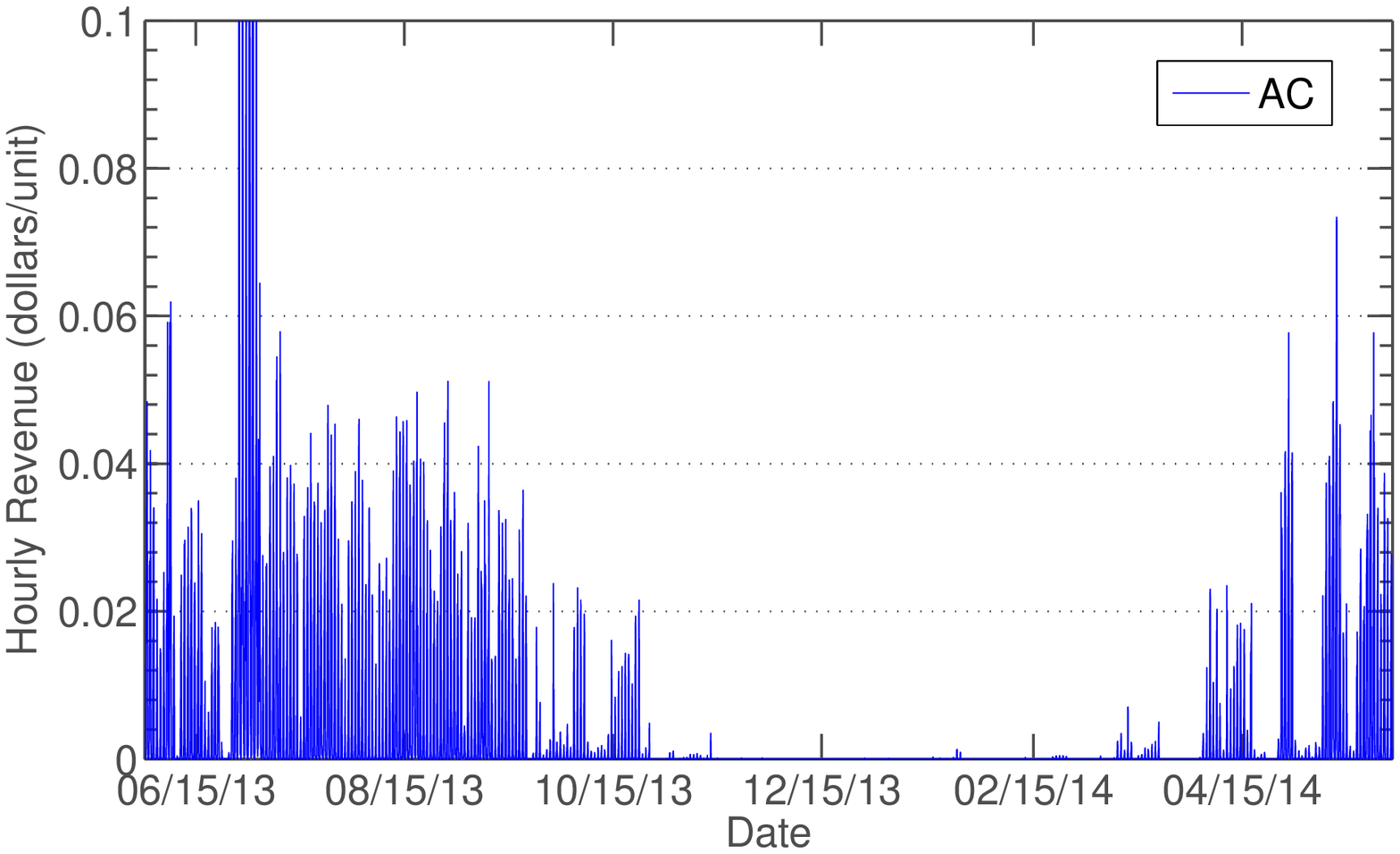}}
\subfigure[Hourly revenue of heat pumps]{\includegraphics[width=.49\columnwidth]{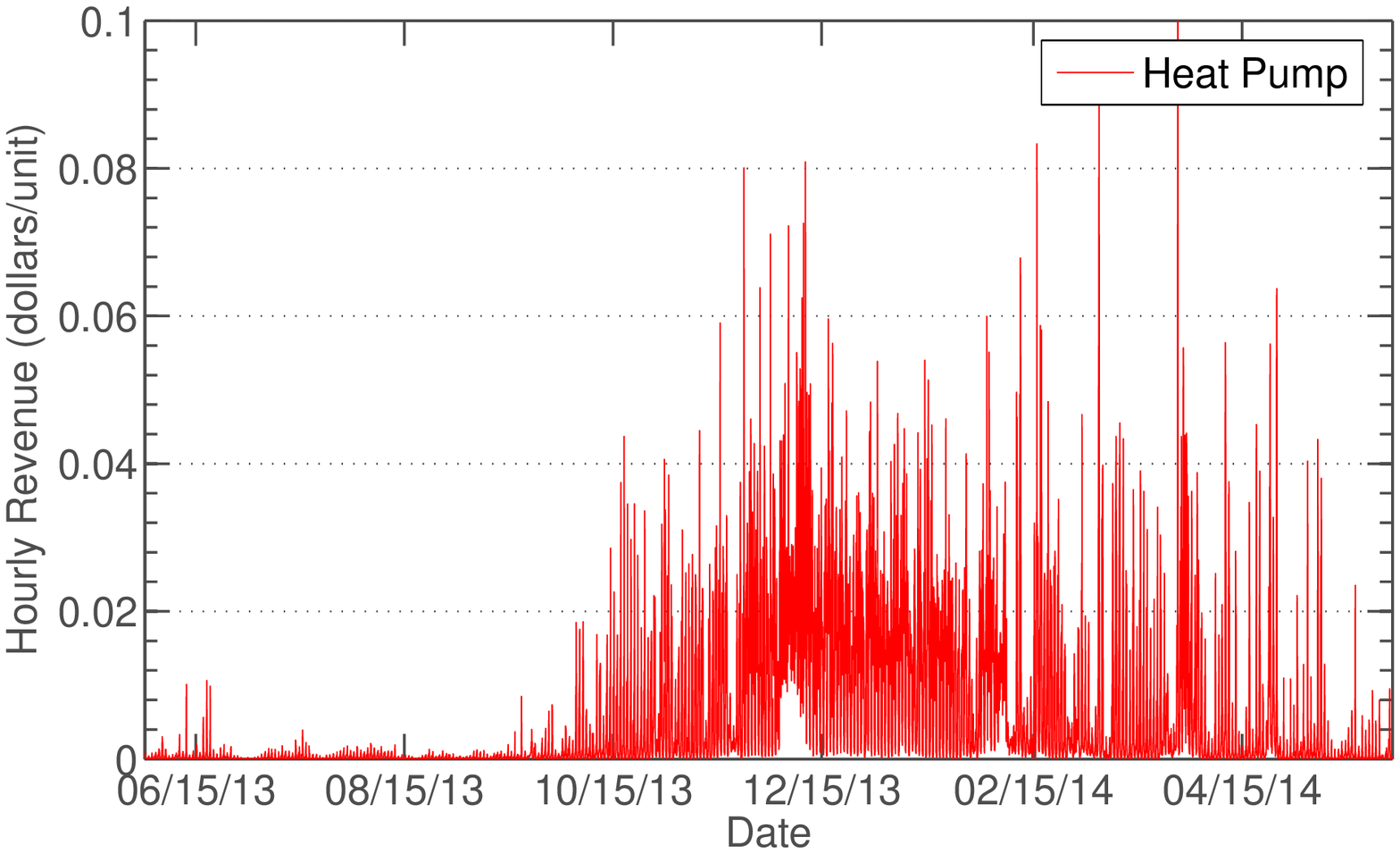}}
\caption{Hourly revenue of ACs (a) and heat pumps (b) in California using the annual temperature profile in Sacramento from June 2013 to May 2014. }\label{fig:yearly_revenue}
\end{figure}

\begin{table}[tb]
\caption{Revenue of \acp{TCL} for providing regulation service in different cities (the unit is $\$$/unit/year). }
\label{tab:revenue}
\vspace{6pt}
\centering
{\scriptsize
\begin{tabular}{c c|| c| c | c | c | c}
 & & SA & SF & BF & LA & SD \\
 \hline \hline
\multirow{3}{*}{AC} & Regulation up capacity & 17.06   & 0.50  & 26.41  & 4.56 & 1.57 \\
 & Regulation up milage  &  0.03  & 0.00  & 0.10  & 0.03  &  0.00 \\
 & Regulation down capacity  & 13.66   & 1.21 & 24.19 & 9.26 & 3.11 \\
 &  Regulation up milage & 0.37 & 0.01 & 1.69 & 0.56 & 0.10 \\
 &  \textbf{Total} & \textbf{31.12}  & \textbf{1.72} & \textbf{52.38} & \textbf{14.41} & \textbf{4.78}  \\
\hline
\multirow{3}{*}{Heat Pump} & Regulation up capacity & 15.40  & 6.51 & 8.78 & 2.36 &  4.12\\
 & Regulation up milage  & 1.02 &  0.44 &  0.55 &  0.15 & 0.28  \\
 & Regulation down capacity  &  26.34 & 18.53 & 18.15  & 7.89 & 12.01 \\
 &  Regulation up milage & 3.76 & 2.34 & 2.68 & 1.03 & 1.41 \\
 &   \textbf{Total} & \textbf{46.52} & \textbf{27.82} & \textbf{30.16} & \textbf{11.43} & \textbf{17.82}   \\
\hline
\multirow{3}{*}{Water Heater} & Regulation up capacity & 9.60  & 9.60  &9.60  &9.60  &9.60    \\
 & Regulation up milage  &0.29 & 0.29 & 0.29 & 0.29 & 0.29  \\
 & Regulation down capacity  & 128.06  & 128.06  & 128.06 & 128.06  & 128.06    \\
 & Regulation up milage & 13.81 & 13.81 & 13.81 & 13.81 & 13.81   \\
 &  \textbf{Total} & \textbf{151.76}  & \textbf{151.76} & \textbf{151.76} & \textbf{151.76} & \textbf{151.76 }   \\
\hline
\multirow{3}{*}{Refrigerator} & Regulation up capacity & 3.93  & 3.93  & 3.93  & 3.93  & 3.93    \\
 & Regulation up milage  &  0.12  &  0.12  &  0.12  &  0.12  &  0.12  \\
 & Regulation down capacity  & 6.09  & 6.09  & 6.09  & 6.09  & 6.09   \\
 & Regulation up milage & 0.66 & 0.66 & 0.66 & 0.66 & 0.66   \\
 & \textbf{Total} & \textbf{10.80}  & \textbf{10.80}  & \textbf{10.80}  & \textbf{10.80}  & \textbf{10.80}    \\
\hline
\end{tabular}}
\end{table}

The regulation revenue consists of two parts: capacity and mileage. It is the summation of the payment from providing reserved capacity and the payment from how many mileage is provided \cite{CAISO_755}. Formally, the revenue is equal to 
\begin{align*}
\text{regulation revenue} = &\text{ regulation capacity price} \times \text{awarded regulation capacity} \\
                + &\text{ regulation mileage price} \times \text{awarded regulation mileage}\times \text{accuracy}.
\end{align*}
Note that the upward regulation and downward regulation are calculated separately. Since the dispatched milage is fixed over a period (e.g., $15$ minutes), higher tracking accuracy will result in higher mileage, and thus larger revenue. In this paper, we assume a tracking accuracy of $95\%$, which is justified by the simulation results reported in our previous work \cite{HH_BS_KP_TV_ACC:2014}. To estimate the annual revenue for regulation service provision, we assume for ACs and heat pumps, the participation functions are as those depicted in Fig.  2. In addition, we assume water heaters and refrigerators are always participating. The hourly revenue is calculated using the 8760-hour regulation price and ambient temperature. 

We estimate the potential revenue for each type of TCLs using the weighted hourly regulation capacity calculated in the last section. The hourly revenue per unit for the four types of TCLs are depicted in Fig.  7. In addition, we also estimate the hourly revenue using the regulation capacity that is calculated based on the temperature profile in Sacramento. The corresponding hourly regulation revenues for ACs and heat pumps are respectively shown in Fig.  8 (a) and Fig.  8 (b). The hourly revenues are more concentrated and higher than those shown in Fig.  7 (a) and Fig.  7 (b). Since the ambient temperatures of water heaters and refrigerators are assumed to be fixed regardless of its residing temperature zone, the estimated hourly regulation revenues of water heaters and refrigerators in this case are the same as those shown in Fig.  7 (c) and Fig.  7 (d).

The annual revenue  (in terms of dollars per unit per year) for each type of \acp{TCL} is summarized in Table IV. We observe that water heaters have the largest potential revenue, while refrigerators have smaller potential revenue due to their smaller rated power (see Table I). In addition, the potential revenue of ACs and heat pumps are highly sensitive to weather. The reason why ACs in San Francisco (SF), Los Angeles (LA), and San Diego (SD) have small revenue potential is because the weather in these regions is moderately cool, which means there is low participation of ACs throughout the year. However, in the cities of Sacramento (SA) and Bakersfield (BF) where the summer is hot, ACs present a large revenue potential. Moreover, if a unit has combined AC and heat pump, and it is located in a region with hot summer and cold winter (e.g., SA and BF), the potential revenue will be substantial.  Additionally, compared to the results reported in \cite{HH_BS_KP_TV_ACC:2014} which only considers ``pay-by-capacity", our revenue estimation is similar to that in \cite{HH_BS_KP_TV_ACC:2014}, although our considered scheme has an additional payment from mileage product. This is mainly because the average MCPs of CAISO for regulation up/down  capacity payment has dropped from $\$5.65/\$4.39$ per MW in 2012 to $\$4.61/\$3.43$ per MW for the period from June 2013 to May 2014 \cite{OASIS}.

In this paper, we have focused our analysis on the CAISO regulation market, whose regulation MCP is relatively low. It is worth to mention that the potential revenue of TCLs providing frequency regulation service is much higher in ISOs such as PJM (Pennsylvania-New Jersey-Maryland Interconnection) and ERCOT (Electric Reliability Council of Texas), whose regulation MCPs are about 2-3 times of that of CAISO \cite{ERCOT_regulation, PJM_price}. A detailed comparison of the regulation market opportunities among different ISOs in the United States is a piece of future work.

\section{Discussion}
\label{sec:econo}
In this section, we estimate the capital cost of instrumentation for \acp{TCL} to provide regulation service, discuss customer incentive methods and qualification requirements, and compare the thermal storage capability of \acp{TCL} with other energy storage technologies. 

\subsection{Capital Cost of \acp{TCL} for Regulation Provision}\label{sec:cost}
The major capital cost enabling \acp{TCL} to provide regulation service consists of 1) meters for real-time power measurement of \acp{TCL}, 2) control devices that override  \acp{TCL}' local control actions,   and 3) communication and control infrastructure that supports metering and telemetry, and can be integrated with \ac{CAISO}'s energy management system. \cut{ In particular, the design, acquisition, and installation of the \ac{CAISO}-approved communication and control equipment shall be under the control of the \ac{CAISO}. }

Measuring the power consumption of each \ac{TCL} necessitates a nontrivial capital cost. In our view, this is unavoidable. Other schemes have been proposed where the aggregate power is {\em estimated} using population models, or {\em disaggregated} from substation measurements \cite{callaway2009tapping, mathieu2013state}. These strategies face the challenges in meeting CAISO's current auditing, telemetry and metering requirements necessary to participate in the regulation market. Moreover,  it is challenging to determine the actual tracking accuracy, and how to settle the payment under the current ``pay-for-performance" scheme becomes a challenge.

For our purpose, we need real-time power measurement of each \ac{TCL} unit every $4$ seconds, in order to satisfy \ac{CAISO}'s telemetry requirement. A smart plug that measures the real-time power consumption of an appliance is about $\$50$ \cite{smart_plug}. Additionally, the price for a smart thermostat that could override a unit's local control actions is about $\$100$ \cite{smart_thermostat}. Regarding communication and control devices, various commercial Remote Telemetry Units (RTUs) and Remote Intelligent Gateways (RIGs) that support internet, cellular and other types of communications are available \cite{kiliccote2010open,brooks2002vehicle,alcoa}. In particular, CAISO's certified metering and telemetry solution providers can be found in \cite{metering_telemetry}. For the purpose of analysis, the combined capital cost enabling ACs and heat pumps to provide regulation service is estimated to be in the range of $100-250$ dollars per unit, and the capital cost for water heaters and refrigerators is estimated to be in the range of $50-100$ dollars per unit.

We would like to point that for some utility companies in California, the basic infrastructure that enables TCLs to provide regulation service is already in place, if the current telemetry and metering requirements for regulation are changed. This would result in reduced cost for TCL power measurement.  As of March 31, 2013, PG$\&$E (Pacific Gas and Electric Company)  has installed over $9$ millions smart meters throughout its service area (covering over $90$ percent of its costumers)  \cite{smart_meter_pge}. The cost of a smart meter that is being used by PG$\&$E is around $\$200$  \cite{smart_meter, smart_meter_pge}. The smart meter data can then be combined with validated HAN (Home and Business Area Network) device to obtain near real-time whole-house level power measurement  \cite{smart_meter_pge}. With smart meters and HAN devices, various non-intrusive load monitoring techniques can be applied to estimate the TCL power \cite{berges2010enhancing}. Additionally, the SmartAC$\texttrademark$ program of  PG$\&$E gathered $147, 600$ residential customers for peak load shaving and managing emergency situations by remotely controlling their AC units \cite{SmartAC}. Such programs can be upgraded to provide ancillary service such as regulation with low additional capital cost. Moreover, we recommend the power measurement, external control, and communication capabilities to be integrated into appliance standards to enable residential \acp{TCL} to provide grid services. This will substantially reduce the capital cost of \acp{TCL} to provide  regulation reserve, and encourage more customers to participate in the demand response program.

\begin{table*}
\caption{Comparison of different energy storage technologies for fast frequency regulation. RU and RD respectively represent regulation up and regulation down.}
\label{tab:comparison}
\vspace{6pt}
{\scriptsize
\begin{tabular}{p{1.115in}||p{0.5in}p{0.7in}p{0.565in}p{0.72in}p{0.93in}p{0.82in}}
Technologies & Technology Maturity   & Cycles/year  & Round-Trip Efficiency  & Cost (\$/kWh) &  \multicolumn{2}{c}{Cost (\$/kW)}  \\\hline \hline
Flywheel   \cite{rastler2010electricity}  & Demo  & $>$8,000 & 85-87\%  & 7,800-8,800 & \multicolumn{2}{c}{1,950-2,200}  \\ 
 Li-ion     \cite{rastler2010electricity}     & Demo   & $>$8,000 & 87-92\%  & 4,340-6,200 & \multicolumn{2}{c}{1,085-1,550}  \\
Advanced lead acid  \cite{rastler2010electricity} & Demo   & $>$8,000 & 75-90\%  & 2,770-3,800 & \multicolumn{2}{c}{950-1,590} \\
Zinc bromide     \cite{sandia_storage}     & Demo    & 5000 & 60-65\%  & 1,464 & \multicolumn{2}{c}{1,464} \\
\hline
AC (SA)    &  R\&D  &  2-3$ \times$nominal & 100\%  &  3,288-8,219 & 494-1,234 (RU) & 209-522 (RD)\\
\hspace{11.5pt} (SF)    &    &   &   &  57,880-144,700 & 14,464-36,160 (RU) & 3,136-7,841 (RD)\\
\hspace{11.5pt}  (BF)   &    &   &   &  1,766-4,415 & 244-611 (RU) & 116-291 (RD)\\
\hspace{11.5pt}  (LA)   &   &   &   &  6,167- 15,415 & 1,242-3,106 (RU) & 353-883 (RD)\\
\hspace{11.5pt} (SD)  &    &   &  &  22,594 -56,485 &  4,395-10,988 (RU) & 1,307-3,267 (RD)\\
Heat Pump   (SA)       &   R\&D & 2-3$ \times$nominal  & 100\%  &  2,395-5,988 & 229-573 (RU)  & 115-286 (RD)\\
\hspace{40pt}   (SF)       &    &   &  &  3,849-9,624 & 514-1,285 (RU)  & 161-403 (RD)\\
\hspace{40pt}   (BF)       &    &  &   &  3,765-9,414 & 400-1,000 (RU)  & 171-429 (RD)\\
\hspace{40pt}   (LA)       &    &   &   &  9,437-23,593 & 1,470-3,675 (RU)  & 378-945 (RD)\\
\hspace{40pt}   (SD)       &   &   &   &  5,802-14,505 & 809-2,022 (RU)  & 240-599 (RD)\\
Refrigerator          &  R\&D  &  2-3$ \times$nominal & 100\%  &  222-444 & 514-1,029 (RU) & 247-493 (RD)\\
Water Heater        &  R\&D  &  4-6$ \times$nominal & 100\%  &83-167 & 211-421 (RU)  &12-23  (RD) 
\end{tabular}
}
\end{table*}

\subsection{Incentive Methods}
Existing demand response incentive methods offer limited financial value. For example, the SmartAC$\texttrademark$ program of PG$\&$E offers a one-time signup bonus of $\$50$ to the customers \cite{SmartAC}.  The OnCall$^\text{\textregistered}$ program of Florida Power and Light Company (FPL) provides $\$5$/month incentive to participating customers for only $7$ months in a year  \cite{OnCall}. As an aggregator, how to incentivize the customers to encourage participation with small reward is a challenge. 

As we showed in the previous section, although aggregation of a collection of \acp{TCL} presents a large potential for regulation provision, the annual revenue  per unit is not substantially attractive if the total revenue is split evenly. Therefore, it is essential to study innovative methods to incentivize residential customers to encourage participation. Studies show that customers prefer lower probability large reward over a guaranteed small reward \cite{kahneman1979prospect}. A lottery-based incentive method has been shown to be very successful in reducing traffic congestion \cite{merugu2009incentive}. It is to our view that incentivizing methods such as lottery-based \cite{merugu2009incentive} or coupon-based  \cite{zhong2013coupon} incentive present a great potential to increase the participation of residential customers for demand response.

\subsection{Comparison with other Energy Storage Technologies}\label{sec:comparison}
In this section, we compare TCLs (thermal storage) with other energy storage technologies that are suitable for providing fast frequency regulation, such as Flywheels, Li-ion, advanced lead acid,  and Zinc Bromide batteries.  According to \cite{rastler2010electricity}, general energy storage application requirements for regulation provision demand the resource to have a size between $1$ MW and $100$ MW with duration of $15$ minutes, have  discharging/charging cycles of more than $8000$ times a year, and have a lifetime of $15$ years. 

The life expectancy for the four types of \acp{TCL} are respectively 15 years for ACs and heat pumps, 14-17 years for refrigerators, and 14 years for water heaters \cite{TCL_life}, which satisfy the application requirements. The cost per kW (respectively kWh) for TCLs are calculated by dividing their capital cost of instrumentation ($\$100-\$250$ for ACs and heat pumps, and $\$50-\$100$ for water heaters and refrigerators) by their average power limits (respectively energy capacity) over the 8760-hour period. Table V shows the comparison results between \acp{TCL} and other storage technologies. Although the storage technologies of flywheel, Li-ion, advanced lead acid, and  Zinc Bromide batteries are more mature,  \acp{TCL} have more competing advantages in terms of cycles, energy efficiency, and cost. In particular, for a collection of \acp{TCL}, provision of regulation service only increases their nominal ON/OFF cycles a few times, unlike flywheels and batteries whose discharging/charging cycles are more than $8000$ times a year. Additionally, the efficiency of \acp{TCL} is much better, which is nearly perfect. Moreover, the costs for water heater and refrigerators are much lower than other technologies. However, we comment that the estimated costs for ACs and heat pumps strongly depend on the temperature profiles used. In cities such as SA and BF, the cost (\$/kW and \$ kWh) for ACs and heat pumps is much lower, while the cost is much higher for TCLs in SF, LA, and SD. Therefore, optimal selection of residential customers for regulation service is an important future work. Furthermore, the disposal of batteries after their retirement will incur additional cost, which otherwise will bring detrimentally environmental effects. This is not the case for TCLs.

\subsection{Qualification Requirements}\label{sec:qualification}
To qualify for regulation service in \ac{CAISO}, a regulating resource must meet the following requirements.

\paragraph{Telemetry and metering} Telemetry refers to real-time measurement data that is sent to the ISO for operational visibility, and metering refers to revenue metering sent to the ISO for settlement purpose.  In \ac{CAISO}, the sampling rate of telemetry is every $4$ seconds. The 1) maximum and minimum operating limits (MW), 2) instantaneous resource output (MW), 3) charge and discharge ramp rates (MW/min), 4) the SoC of the NGR, 5) connectivity status (ON/OFF), and 6) Automatic Generation Control (AGC) Status (Remote/Local), are sent to the ISO's energy management system every $4$ seconds. More detailed requirements on telemetry data point for NGRs can be found in \cite{CAISO_Telemetry}. \ac{CAISO} requires the metering accuracy to be $\pm 0.25\%$. Moreover, the data should be directly measured from the resource instead of an aggregation. These requirements impose a non-trivial cost on power measurement and communication on each \ac{TCL}.

\paragraph{Minimum available power} \ac{CAISO} requires the regulation resource to have at least $0.5$ MW power capacity to be able to participate in the frequency regulation market. This corresponds to a collection of about $200$ available ACs, heat pumps, or water heaters, or a population of about $3500$ refrigerators. However, aggregation of DERs is not allowed in \ac{CAISO}'s ancillary service market. In fact, among all ISOs/RTOs in the United States, aggregation is only allowed in \ac{PJM} subject to approval.

\paragraph{Minimum continuous energy delivery time} In \ac{CAISO}, the minimum continuous energy delivery time in the day-ahead and real-time market are respectively $60$ and $30$ minutes. However, for NGR with REM, the resources in \ac{CAISO} are allowed to bid their $15$-min energy capacities into the regulation market and adjust their SoCs to the desired values in the real-time energy market. This scheme takes advantage of the fast ramping potential of regulation resources such as \acp{TCL} and allow those resources with limited energy capacities to be able to bid larger capacity and achieve more revenue in the regulation market than the case without REM. 

\paragraph{Minimum performance threshold} CAISO requires regulating resources to have at least $50\%$ tracking accuracy for both regulation up and regulation down measured over a calendar month. Re-certification is required within $90$ days if the resource fails the minimum performance requirement. The accuracy performance in CAISO is measured by the ratio of the sum of the AGC (Automatic Generation Control) setpoint for each 4-second regulation interval less the sum of  the tracking error for each regulation interval to the sum of the AGC setpoint. The accuracy percentage is calculated every $15$ minutes.  In \cite{HH_BS_KP_TV_ACC:2014}, it was shown that even with up to $20$-second communication delay, the tracking accuracy of \acp{TCL} could easily satisfy the CAISO accuracy performance threshold. \cut{ In \ac{PJM}, the participating resources are required to pass several tests with performance scores of $0.75$ or above in order to be qualified for regulation service.}

\paragraph{Ramping rate requirement} The participating resources are required to ramp to their  awarded capacity in $10$ minutes.  This requirement can also be easily satisfied by an aggregation of \acp{TCL}. In fact, one of the main advantages of a large collection of \acp{TCL} is their fast aggregate ramping rate; this is because \acp{TCL} can be turned ON or OFF simultaneously. Moreover, fast-ramping resources are able to provide more mileage when providing regulation service, and make more revenue than slow-ramping thermal generators. 

\paragraph{Others} To participate as NGR resource, \ac{CAISO} will require resources to sign participating agreements, conform to resource data template, and use existing business processes. In addition, all market participants must be represented by a Scheduling Coordinator that is financially responsible for all interactions with the market and undergoes special certification. Moreover, CAISO requires that the scheduling coordinator must be a load serving entity.

\section{Conclusions and Policy Implications}
\label{sec:conclusion}

\subsection{Conclusions and Future Work}
We presented the potentials and economics of using \acp{TCL} to provide frequency regulation service to the grid. In particular, we estimated the potential resource, capital cost, and revenue of \acp{TCL}, discussed the incentive methods and qualification requirements, and compared \acp{TCL} with other energy storage technologies. Our results showed that the potential of \acp{TCL} in California was more than enough for provision of regulation service for now and the near future. Moreover, we showed  that the cost of instrumentation for \acp{TCL} providing regulation reserve was more favorable than flywheels, Li-ion, advanced lead acid, and Zinc Bromide batteries. 

There are several important future research issues that must be addressed. These include: (a) characterizing battery model parameters that account for uncertainties of TCL model and ambient temperature, (b) estimating the overall hourly availability of TCLs using historic measurement data, (c) conducting pilot programs to showcase the feasibility of using TCLs to provide regulation reserve, (d) developing fair schemes to incentivize \ac{TCL} owners to participate in frequency regulation service, and (e) optimal targeting of residential customers for 
demand response. 

\subsection{Recommended Policy Changes}
In this section, we present some recommended policy changes to enable \acp{TCL} to participate in the \ac{CAISO} regulation market. 

Based on the discussion on cost of instrumentation in Section \ref{sec:cost}, we see that the capital cost of retrofitting TCLs to provide regulation service is still relatively high compared to their revenue. We comment that design and adoption of new appliance standards such as adding power measurement, external control, and communication capabilities is desirable to reduce the cost of instrumentation for TCLs. Additionally, allowing to use NIALM (Non-Intrusive Appliance Load Monitoring) technique also has a great potential to reduce the capital cost.  

Besides recommending the above policy changes on allowing new standards and techniques, we also present some recommended regulatory changes on qualification requirements. Recall in Section \ref{sec:qualification} that the telemetry and metering requirements impose a non-trivial cost on power measurement and communication on each \ac{TCL}. We comment that a change on the sampling frequency, aggregation rule, metering accuracy, and enabling low cost communication such as using internet is necessary to lower the capital cost of \acp{TCL} to provide regulation service. In fact, CAISO is currently examining how to expand current metering and telemetry methods  to facilitate market participation by Distributed Energy Resources (DERs) such as dispatchable demand response, energy storage, distributed generation, and NGRs \cite{CAISO_telemetry_change}. Moreover, it is suggested that the current CAISO regulatory rule ``no aggregation of DERs is allowed" discussed in Section \ref{sec:qualification}  also needs to be changed to enable DERs such as TCLs to participate in the CAISO ancillary service market. 

In summary, we comment that certain policies must be changed in order to enable TCLs to participate in the CAISO regulation service market. These include allowing aggregation of DERs, relaxation of telemetry and metering requirements, and design and adoption of new appliance standards and techniques.

\section*{Acknowledgment}
The authors gratefully acknowledge Pravin Varaiya for insightful comments and discussions. This work was supported in part by EPRI and CERTS under sub-award 09-206; PSERC S-52; NSF under Grants CNS-0931748, EECS-1129061, CPS-1239178, and CNS-1239274; the Republic of Singapore National Research Foundation through a grant to the Berkeley Education Alliance for Research in Singapore for the SinBerBEST Program; Robert Bosch LLC through its Bosch Energy Research Network funding program.

\section*{References}
\bibliographystyle{IEEEtran}    
\bibliography{TCL}

\end{document}

%% file: my_preamble.tex
\newtheorem{theorem}{Theorem}

\newtheorem{definition}{Definition}

\acrodef{TCL}{Thermostatically Controlled Load}
\acrodef{AC}{Air Conditioning}
\acrodef{FERC}{Federal Energy Regulatory Commission}
\acrodef{AGC}{Automatic Generation Control }
\acrodef{CAISO}{California Independent System Operator}
\acrodef{PJM}{Pennsylvania-New Jersey-Maryland}
\acrodef{MCP}{Market Clearing Price}
\acrodef{NGR}{Non-Generator Resource}
\acrodef{SoC}{State of Charge}
\acrodef{EV}{Electric Vehicle}
\acrodef{RES}{Renewable Energy Resource}
\acrodef{CAISO}{California Independent System Operator}